\documentclass[graybox]{svmult}

\usepackage{mathptmx}         
\usepackage{graphicx}         
\usepackage{longtable}         

\newcommand{\rmn}
  {\rm}
\newcommand{\abs}
  {{\rm abs}}

\newlength{\doublefigwidth}
\newlength{\figwidth}
\newlength{\thinfigwidth}
\newcommand{\unit}[1]
  {{\mbox{\rm\,\,#1}}}
\newcommand{\percent}
  {\%}
\newcommand{\lya}
  {Ly\,$\alpha$}
\newcommand{\lyb}
  {Ly\,$\beta$}
\newcommand{\lyg}
  {Ly\,$\gamma$}
\newcommand{\lyd}
  {Ly\,$\delta$}
\newcommand{\nv}
  {N\,{\sc{v}}}
\newcommand{\mgii}
  {Mg\,{\sc{ii}}}
\newcommand{\mabs}
  {{M_{1450}}}

\newcommand{\lsun}
  {L_{\odot}}

\newcommand{\hzq}
  {HZQ}
\newcommand{\hzqs}
  {HZQs}

\newcommand{\hi}
  {{\textrm{H\,{\sc{i}}}}}

\newcommand{\hii}
  {{\textrm{H\,{\sc{ii}}}}}

\newcommand{\nir}
  {NIR}

\newcommand{\redshift}
  {z}

\newcommand{\rest}
  {{\rm em}}

\newcommand{\etc}
  {etc.}
\newcommand{\etal}
  {et al.}
\newcommand{\eg}
  {e.g.}
\newcommand{\cf}
  {cf.}
\newcommand{\ie}
  {i.e.}
\newcommand{\diff}
  {{\rmn{d}}}

\newcommand{\eq}[1]
  {Eq.~\ref{equation:#1}}

\newcommand{\Eq}[1]
  {Equation~\ref{equation:#1}}

\newcommand{\sect}[1]
  {Section~\ref{section:#1}}

\newcommand{\tabl}[1]
  {{\mbox Table~\ref{table:#1}}}

\newcommand{\fig}[1]
  {Fig.~\ref{figure:#1}}

\newcommand{\Fig}[1]
  {Figure~\ref{figure:#1}}

\newcommand{\step}
  {\Theta}
\newcommand{\la}
  {\ {\raise-.5ex\hbox{$\buildrel<\over\sim$}}\ }
\newcommand{\ga}
  {\ {\raise-.5ex\hbox{$\buildrel>\over\sim$}}\ }
\newcommand{\micron}
  {$\mu$m}

\newcommand{\age}
  {T_{\rm{q}}}
\newcommand{\rateion}
  {\Gamma_{\rm{ion}}}
\newcommand{\nhi}
  {\bar{n}_{{\rm H\,I}}}
\newcommand{\nhilos}
  {n_{{\rm H\,I}}}

\newcommand{\rnz}
  {R_{\rm NZ}}
\newcommand{\rhii}
  {R_{\rm H\,II}}

\newcommand{\rnzcorr}
  {R_{\rm NZ,corr}}

\newcommand{\fhi}
  {x_{{\rm H\,I}}}
\newcommand{\subhi}
  {{\rm H\,I}}
\newcommand{\subhii}
  {{\rm H\,II}}
\newcommand{\lambdaalpha}
  {\lambda_\alpha}

\newcommand{\diracdelta}
  {\delta_{\rm D}}
\newcommand{\src}
  {{\rm src}}
\newcommand{\obs}
  {{\rm obs}}

\newcommand{\gp}
  {{\rm GP}}
\newcommand{\einstein}
  {\Lambda}
\newcommand{\vlos}
  {v_{||}}
\newcommand{\sigmalos}
  {\sigma_{||}}
\newcommand{\start}
  {{\rm start}}
\newcommand{\finish}
  {{\rm end}}
\newcommand{\nz}
  {{\rm NZ}}

\newcommand{\rydberg}
  {E_\infty}
\newcommand{\mathif}
  {\,\,\,\,\,\,\,\,\,\, {\rm if} \,\,\,\,\,\,\,\,\,\,}

\begin{document}

\title*{Quasars as probes of cosmological reionization}
\titlerunning{Quasars and reionization}
\author{Daniel J.\ Mortlock}
\authorrunning{D.\ J.\ Mortlock}
\institute{Astrophysics Group, 
  Blackett Laboratory, 
  Imperial College London, 
  London SW7 2AZ, UK \\
  \email{mortlock@ic.ac.uk}}

\maketitle

\abstract{Quasars are the most luminous non-transient 
  sources in the epoch of cosmological reionization 
  (\ie, which ended a billion years after the Big Bang,
  corresponding to a redshift of $\redshift \simeq 5$),
  and are powerful probes of the 
  inter-galactic medium at that time.
  This review covers current efforts to 
  identify high-redshift quasars 
  and how they have been used to constrain
  the reionization history.
  This includes a full description of the 
  various processes by which neutral hydrogen atoms
  can absorb/scatter ultraviolet photons,
  and which lead to 
  the Gunn-Peterson effect, 
  dark gap and dark pixel analyses, 
  quasar near zones and 
  damping wing absorption.
  Finally, the future prospects for using quasars as probes of reionization
  are described.}


\section{Introduction}
\label{section:intro}

Cosmological hydrogen reionization was largely complete 
by about a billion years after the Big Bang, 
when the Universe was
just $\sim 7\%$ of its current age ($\sim 13.8\,\unit{Gyr}$).
To learn directly about the process of reionization requires that 
signals -- most obviously photons -- are detected from 
this epoch or earlier (although inferences 
can also be made from sources seen at later times).
The cosmic microwave background (CMB) radiation can be used as 
a ``back-light'' to probe reionization, providing important
integral constraints on the density of free electrons (Chapter~X)
and 21~cm radiation from neutral hydrogren (\hi) 
present in the early Universe should be able to provide a
wealth of information in the future (Chapter~X),
but the most direct probes are the luminous astronomical
objects that had formed before reionization was complete.

\begin{figure}[t]
\begin{center}
\includegraphics[scale=.6]{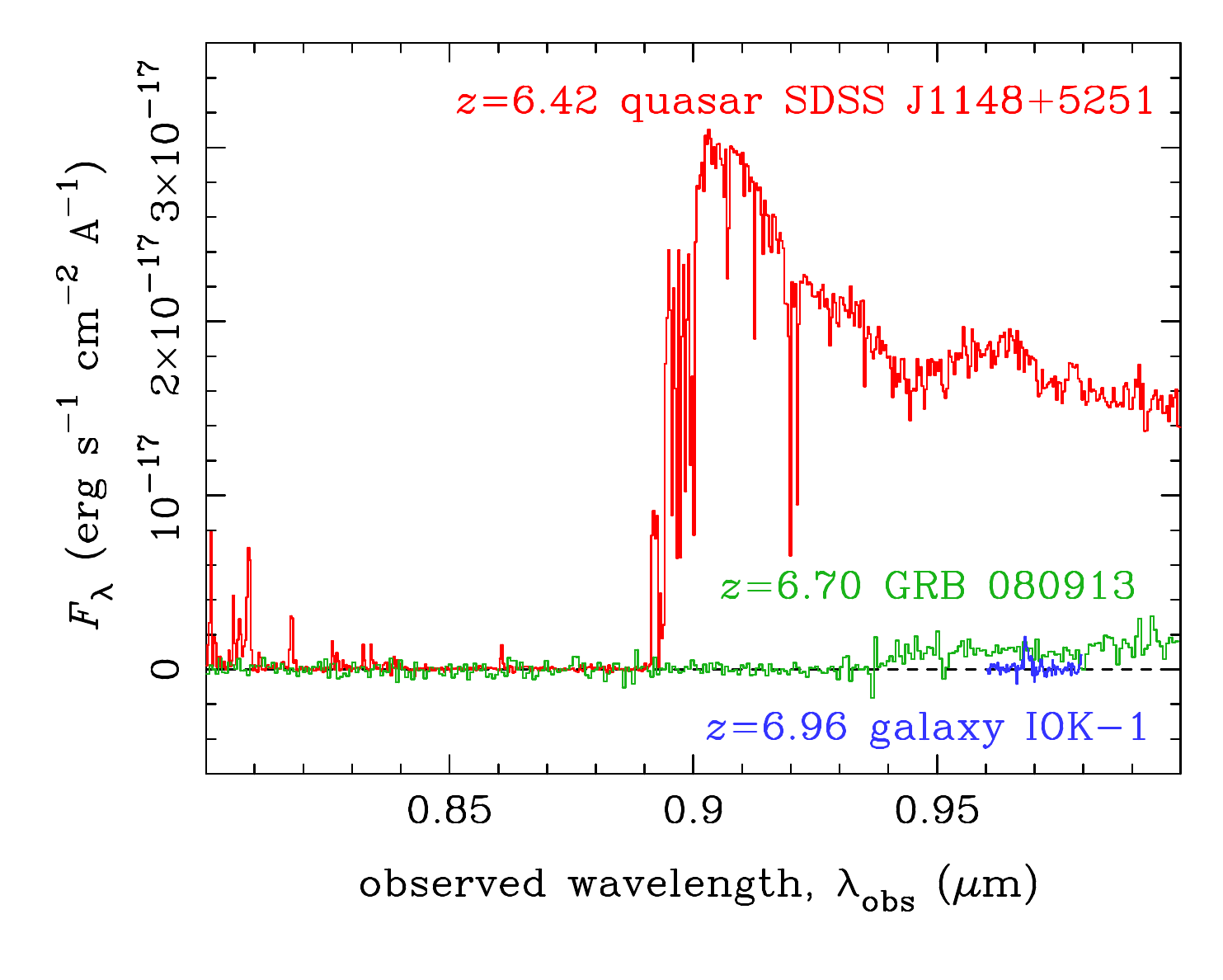}
\caption{Spectra of representative high redshift sources:
  the $\redshift = 6.42$ quasar SDSS~J1148+5251
  \cite{Fan_etal:2003,White_etal:2003};
  GRB~080913 at $\redshift = 6.70$ \cite{Greiner_etal:2009};
  and the \lya-emitting $\redshift = 6.96$ galaxy IOK-1 \cite{Iye_etal:2006}.}
\end{center}
\label{figure:spectra}
\end{figure}

Seen with redshifts of $\redshift \ga 6$, these sources 
are amongst the most distant currently known to astronomy.
Ordinary galaxies at these distances are too faint to provide
much information individually, 
but their properties as a population are sufficiently well 
constrained for them to be confirmed as 
the dominant source of ionizing photons (Chapter~X).
Much brighter 
gamma-ray bursts (GRBs) have also been detected during the reionization
epoch; their largely featureless spectra are ideal for
absorption studies, although a combination of their rarity and transience 
have so far limited their effectiveness as
reionization probes (Chapter~X).
The other main class of astronomical object known to have existed
at these early times are quasars.
They have 
high luminosities (comparable to GRBs at peak and  
thousands of times brighter than field galaxies, as illustrated
in \fig{spectra})
but are extremely rare,
with only a few per thousand deg$^2$ on the sky.

The use of quasars as probes of reionization is explored here,
updating previous reviews 
(\cite{Fan_etal:2006a,Djorgovski_etal:2006} 
and the introductions to, \eg, 
\cite{Fan_etal:2006b,McGreer_etal:2011})
and unifying some of the varied theoretical approaches 
taken in cosmology textbooks 
(\eg, \cite{Peebles:1993,Peacock:1999}). 
The emphasis is mostly on the underlying theory and methods,
as these are most likely to remain robust to future discoveries,
although the more secure results are given as well.
The context is given by the cosmological model
(\sect{cosmology})
and a summary of the results of quasar searches 
(\sect{quasars}).
The absorption properties of neutral hydrogen 
(\sect{los})
lead to a number of distinct constraints on the 
neutral fraction of hydrogen in the high-redshift Universe
(Sections~\ref{section:gp}, \ref{section:nearzone} and \ref{section:wing}).
With new data and better models -- and ways of
comparing the two -- 
these constraints will steadily be improved in the future (\sect{future}).

\section{Cosmology}
\label{section:cosmology}

Cosmological distances, volumes, \etc, 
are calculated assuming that the Universe is spatially flat,
the normalised matter density is 
$\Omega_{\rm m} = 0.3$, 
the normalised cosmological constant is 
$\Omega_{\Lambda} = 1 - \Omega_{\rm m} = 0.7$,
the normalised baryon density is 
$\Omega_{\rm b} = 0.04$,
the primordial helium fraction (by mass) is 
$Y = 0.24$,
and the Hubble constant is
$H_0 = 70 \unit{km} \unit{s}^{-1} \unit{Mpc}^{-1}$.

All lengths, volumes, \etc, are proper, {\em not} co-moving,
with the exception of the mean total density of hydrogen
(\ie, both neutral and ionized),
which is given in terms of the
current (and hence co-moving) value,
\begin{equation}
\bar{n}_{\rm H,0} 
  \, = \,
  \frac{3 \, H_0^2 \, (1 - Y) \, \Omega_{\rm b}}
  {8 \pi \, G \, m_{\rm p}}
  \, \simeq \, 0.17 \unit{atoms} \unit{m}^{-3},
\end{equation}
where
$G = 6.67 \times 10^{-11} \unit{m}^3 \unit{kg}^{-1} \unit{s}^{-2}$
is Newton's gravitational constant
and
$m_{\rm p} = 1.67 \times 10^{-27} \unit{kg}$ is the proton mass.
The mean proper density of neutral hydrogen in the
inter-galactic medium (IGM) at redshift $\redshift$
can hence be written in terms of the
evolving mean neutral fraction,
$\fhi(\redshift)$,
as
\begin{equation}
\label{equation:nhicosmo}
\nhi(\redshift) 
  \, = \, \bar{n}_{\rm H,0} \, (1 + \redshift)^3 \, \fhi(\redshift) 
  \, = \, \frac
  {3 \, H_0^2 \, (1 - Y) \, \Omega_{\rm b} \, (1 + \redshift)^3 \, 
    \fhi(\redshift)}
  {8 \pi \, G \, m_{\rm p}}.
\end{equation}
Deviations from uniformity that arise in particular from 
the inhomogeneous or ``patchy'' nature of reionization
(Chapter~X)
are discussed where relevant.

The Hubble parameter evolves with redshift according to
\begin{equation}
\label{equation:h_z}
H(\redshift) 
  \, = \, H_0 \left[
  \Omega_{\rm m}(1 + \redshift)^3
  + 
  \Omega_{\Lambda}
  - 
  (\Omega_{\rm m} + \Omega_\Lambda - 1) (1 + \redshift)
  \right]^{1/2}
  \simeq H_0 \, \Omega_{\rm m}^{1/2} \, (1 + \redshift)^{3/2},
\end{equation}
where the second expression is accurate to per cent level
for the redshifts of $\redshift \ga 5$ that are relevant here.
In this regime 
a redshift of $\redshift$ corresponds to a time 
(measured since the Big Bang at time $t = 0$) of
\begin{equation}
t \, \simeq \, 
  \frac{2}{3 \, \Omega_{\rm m}^{1/2} \, H_0 \, (1 + \redshift)^{3/2}}
  \simeq 0.92 \, \left(\frac{1 + \redshift}{7}\right)^{-3/2} \unit{Gyr}.
\end{equation}

The change in redshift, $\diff \redshift$, 
that occurs while a photon moves a (proper) distance
$\diff l$ towards Earth
is given implicitly by
\begin{equation}
\label{equation:dl}
\diff l 
\, = \, 
c \, \diff t
\, = \, 
 \frac{- c}{(1 + \redshift) \, H(\redshift)} \, \diff \redshift
\simeq 
  \frac{- c}{H_0 \, \Omega_{\rm m}^{1/2} \, (1 + \redshift)^{5/2}} \,
  \diff \redshift,
\end{equation}
where the last expression is, again, accurate for $\redshift \ga 5$.
This relationship is needed both for optical depth
calculations (\sect{expand}) 
and to convert from small wavelength shifts to local physical
scales (\sect{nearzone}).
In particular, 
a photon 
emitted by a distant source at redshift $\redshift_\src \ga 5$
would reach a small (proper) distance $R$
at a cosmological time corresponding to a redshift of
\begin{equation}
\label{equation:z_r}
\redshift 
  \, \simeq \, 
    \redshift_\src
    - \frac{R}{c / H_0} 
    \Omega_{\rm m}^{1/2} (1 + \redshift_\src)^{5/2},
\end{equation}
where it is assumed that the light travel time,
$R / c$, is much smaller than the Hubble time, $1 / H(\redshift_\src)$,
at that redshift.
If the photon's 
wavelength had been $\lambda$ when it was at distance $R$ 
from the source then
it would now be observed with a wavelength of 
\begin{equation}
\label{equation:lambda_r}
\lambda_\obs \, \simeq \,
  \lambda \, (1 + \redshift_\src)
  \left[ 1 - \frac{R}{c / H_0} 
  \Omega_{\rm m}^{1/2} (1 + \redshift_\src)^{3/2} \right].
\end{equation}


\section{High-redshift quasars}
\label{section:quasars}

Quasars 
were first identified in 1963
\cite{Hazard_etal:1963,Schmidt:1963} as 
bright point-sources with what were then thought of 
as unusually high redshifts. 
The nature of these objects was unknown at that time,
although a consensus has since been reached that a quasar is
the observable manifestation of 
the hot, compressed gas in an accretion disk around 
a super-massive black hole in the centre of an 
otherwise ordinary galaxy (\eg, \cite{Rees:1984}).

The physical nature of quasars is, however, largely irrelevant to their 
utility as probes of reionization; 
more important is that 
bright quasars 
(with bolometric
luminosities of $L \ga 10^{40} \unit{W} \simeq 10^{13} \lsun$) 
had already 
formed by the time the hydrogen in the IGM
was undergoing reionization.
These high-redshift quasars (\hzqs) are seen as unresolved point-sources
with optical or near-infrared (NIR) flux densities of 
$\sim 0.01 \unit{mJy}$, corresponding to 
(AB) magnitudes of $\sim 20$,
although there is
(as discussed further in \sect{gp})
almost complete absorption blueward of the 
\lya\ break (\sect{lyman_series}) at 
an observed wavelength of 
$
\lambda_\obs
   \simeq 
  [0.85 + 0.12 (\redshift - 6)] \unit{\micron}$.
Critically, high quality spectra of $\redshift \ga 6$ quasars 
can be obtained with the current generation of
large (\ie, $8\unit{m}$ class) ground-based telescopes.
Such observations make detailed absorption studies possible,
and most of the reionization constraints discussed below
are based on measurements of this sort.

The practical utility of quasars as probes of reionization is 
limited primarily by the 
fact that they are so rare, 
with a number density of just a few per cubic Gpc at 
$\redshift \simeq 6$ \cite{Fan_etal:2001,Willott_etal:2010},
corresponding to a few per thousand deg$^2$ on the sky.
Fortunately, there are a number of compelling reasons
(including finding more \hzqs!)
to undertake wide-field sky surveys 
(\sect{surveys}),
with the result that 
the number of known \hzqs\ is now close to one hundred (\sect{known}).


\subsection{High-redshift quasar surveys}
\label{section:surveys}

Observational astronomy is increasingly based on large
(\ie, either wide and/or deep) 
sky surveys, 
particularly 
at the optical and NIR wavelengths that can be used 
to identify $\redshift \ga 6$ sources from their \lya\ break.
(It is also possible to identify \hzqs\ by 
exploiting the fact that some are radio-loud 
\cite{McGreer_etal:2006,Cool_etal:2006,Zeimann_etal:2011},
but the optical-NIR search methods have resulted in the vast
majority of discoveries to date.)
The Sloan Digital Sky Survey (SDSS \cite{York_etal:2000}),
the Canada France Hawaii Telescope Legacy Survey
  (CFHTLS \cite{Astier_etal:2006,Gwyn:2012}),
the Panoramic Survey Telescope And Rapid Response System  
  (Pan-STARRS \cite{Kaiser_etal:2002,Hodapp_etal:2004}),
the UKIRT Infrared Deep Sky Survey 
  (UKIDSS \cite{Lawrence_etal:2007})
and
the Visible and Infrared Survey Telescope for Astronomy
  (VISTA \cite{Sutherland_etal:2014})
have, between them, observed over half the
sky and catalogued more than a billion sources
in the optical and NIR.
All these surveys have the combination of area,
depth and wavelength coverage needed to detect significant numbers of
\hzqs,
although it is only the surveys with multi-band NIR coverage
(UKIDSS, VISTA and, to a lesser degree, Pan-STARRS) 
that can probe $\redshift \ga 6.5$.

All the above surveys 
have discovered several \hzqs\ 
(including a number of rediscoveries of previously known
objects, an important cross-check).
While the search process is complicated
(see, \eg, 
  \cite{Fan_etal:2001,Willott_etal:2007,Jiang_etal:2008,Mortlock_etal:2012a}),
there do not appear to be any significant selection 
effects that would bias the inferences about reionization made 
from the sub-set of quasars that have been found.
In particular,
even though the majority of the quasars listed in \tabl{quasars}
were initially discovered by exploiting the fact that 
they have no appreciable flux in all but the reddest optical passbands,
there is no suggestion of a sub-population of $\redshift \ga 6$
sources without sharp \lya\ breaks that have been missed.
The corollary is that any \hzq\ can be used to 
provide constraints on the evolution of $\fhi$, 
although the fact that reionization is expected to be a
patchy process (Chapter~X) means that 
that multiple lines-of-sight are needed to 
extrapolate to the global reionization history.


\subsection{Currently known high-redshift quasars}
\label{section:known}

\begin{figure}[t]
\begin{center}
\includegraphics[scale=.6]{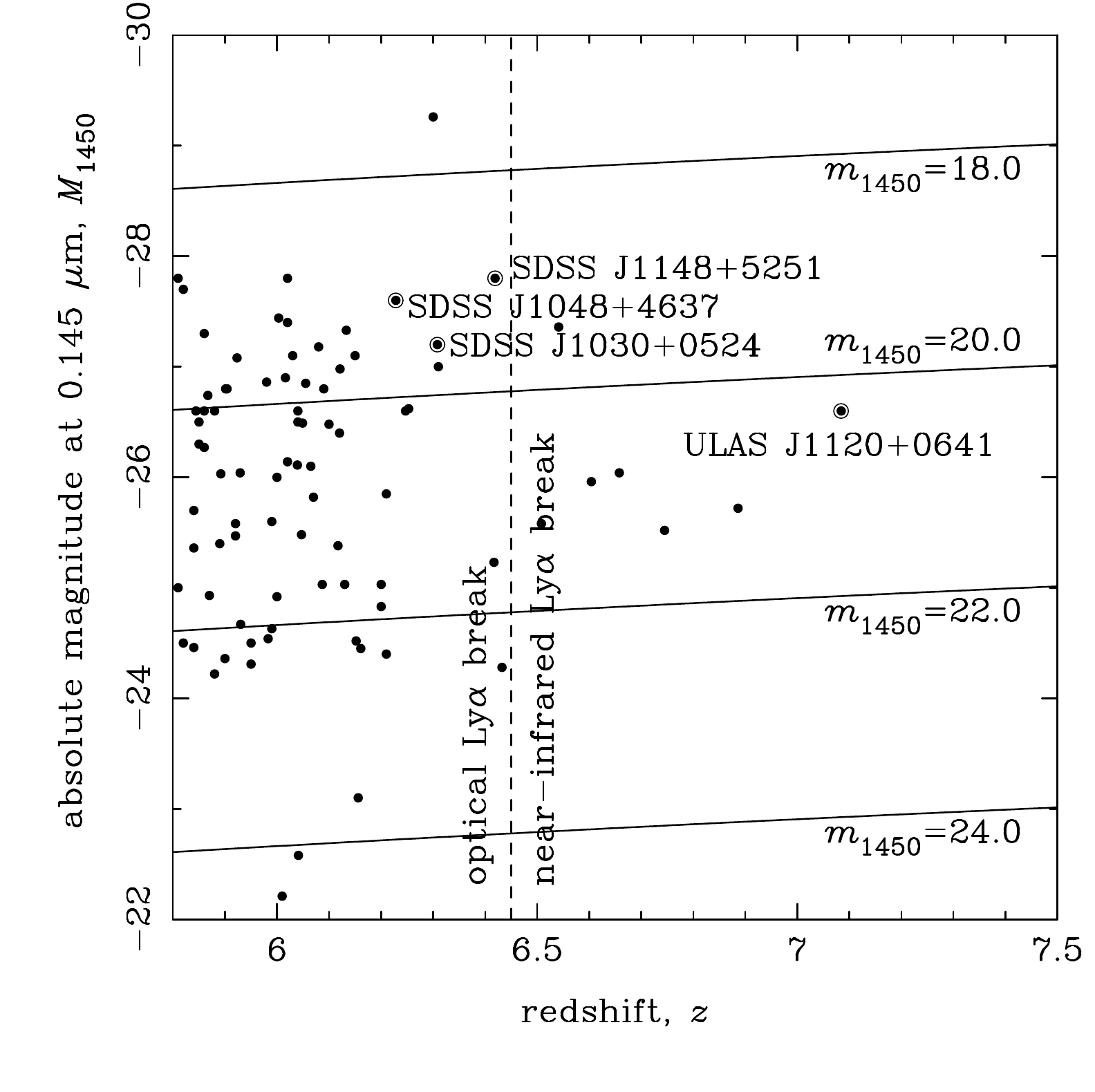}
\caption{Redshifts and absolute magnitudes (measured at a rest-frame
wavelength of $\lambda_\rest = 0.145 \unit{micron}$) of the known 
$\redshift \geq 5.8$ quasars
listed in \tabl{quasars}.
The four quasars discussed in \sect{known} are labelled explicitly.
The solid curves trace lines of
constant AB magnitude, $m_{1450}$,
\ie, the flux density at an observed wavelength of
$\lambda_\obs = 0.145 \, (1 + \redshift) \unit{\micron}$.
The dashed vertical line indicates the redshift above which,
due to \lya\ absorption by the IGM,
sources are detectable only at \nir\ wavelengths.}
\end{center}
\label{figure:quasarplot}
\end{figure}

The above surveys -- and, to a lesser degree, other searches --
have,
as of mid-2015, resulted in the discovery of 89 quasars with redshifts
of $\redshift \geq 5.8$.
These are listed in \tabl{quasars}\footnote{An
up-to-date and expanded version
of this table in machine-readable form
is available from the author.}
and illustrated in Fig.~2.

For reionization studies it is the
most distant and the most luminous\footnote{Strictly, 
it is the flux at Earth which 
is the relevant quantity but, as 
can be seen from Fig.~2, the decrease in flux with
redshift across the reionization epoch 
is considerably smaller than the range of quasar luminosities,
so $\mabs$ is an excellent proxy for the likely utility of any single
source.} which are most important.
The \hzqs\ which have been subject to the most scrutiny so far are:
\begin{itemize}
\item
SDSS~J1030+0524 \cite{Fan_etal:2001}
  at $\redshift = 6.30$, 
  which was the most distant known quasar when it was 
  discovered and is currently the 
  third brightest with $\redshift \geq 6$;
\item
SDSS~J1048+4637 \cite{Fan_etal:2003}
  at $\redshift = 6.26$,
  which is currently the second brightest known quasar
  with $\redshift \geq 6$;
\item
SDSS~J1148+5251 \cite{Fan_etal:2003} 
  at $\redshift = 6.42$, 
  which was also the most distant known quasar when it was
  discovered
  and is currently the brightest known with $\redshift \geq 6$;
\item
ULAS~J1120+0641 \cite{Mortlock_etal:2011d} at 
  $\redshift = 7.08$, 
  which is currently the most distant quasar known.
\end{itemize}
These four\footnote{The recently discovered quasars
PSO~J0226+0302 \cite{Venemans_etal:2015},
at $\redshift = 6.53$ and unusually bright,
and 
SDSS~J0100+2802 \cite{Wu_etal:2015}, 
at $\redshift = 6.30$ and a factor of a few
more luminous than any other known \hzq,
will presumably 
be key sources in the future, but as yet have not been subject 
to full follow-up campaigns.}
objects are highlighted in Fig.~2,
indicating why SDSS~J1148+5251 and 
ULAS~J1120+0641 in particular are central to a number of the 
reionization constraints discussed below.

{\footnotesize
\begin{longtable}{rcccc}
\label{table:quasars} \\
\multicolumn{5}{l}{{\bf \tablename\ \thetable{}} Known high-redshift ($\redshift \geq 5.8$) quasars} \\
\hline
\multicolumn{1}{c}{ID}&\multicolumn{1}{c}{$\redshift$}&\multicolumn{1}{c}{$M_{1450}$}&\multicolumn{1}{c}{$\redshift_{\rm NZ}$}&\multicolumn{1}{c}{reference(s)}\\
\hline
\endfirsthead
\multicolumn{5}{l}
{{\bf \tablename\ \thetable{}} Continued from previous page} \\
\hline
\multicolumn{1}{c}{ID}&\multicolumn{1}{c}{$\redshift$}&\multicolumn{1}{c}{$M_{1450}$}&\multicolumn{1}{c}{$\redshift_{\rm NZ}$}&\multicolumn{1}{c}{reference(s)}\\
\hline
\endhead
\hline \multicolumn{5}{r}{{Continued on next page}} \\
\endfoot
\hline
\endlastfoot
SDSS J000239.39$+$255034.8 & $5.820\pm0.020$ & $-27.70\pm0.00$ & $5.65\pm0.01$ & \cite{Fan_etal:2004,Carilli_etal:2010}\\
SDSS J000552.34$-$000655.8 & $5.850\pm0.003$ & $-26.50\pm0.10$ & $5.77\pm0.01$ & \cite{Fan_etal:2004,Kurk_etal:2007,Carilli_etal:2010}\\
SDSS J000825.77$-$062604.6 & $5.929\pm0.003$ & $-26.04\pm0.09$ & ... & \cite{Jiang_etal:2015}\\
PSO J002806.56$+$045725.7 & $6.040\pm0.030$ & $-26.50\pm0.10$ & ... & \cite{Banados_etal:2014,Jiang_etal:2015}\\
CFHQS J003311.40$-$012524.9 & $6.130\pm0.020$ & $-25.03\pm0.10$ & ... & \cite{Willott_etal:2007}\\
CFHQS J005006.67$+$344522.6 & $6.253\pm0.003$ & $-26.62\pm0.10$ & ... & \cite{Willott_etal:2010a,Willott_etal:2010b}\\
CFHQS J005502.91$+$014618.3 & $5.983\pm0.004$ & $-24.54\pm0.10$ & ... & \cite{Willott_etal:2009,Willott_etal:2010a}\\
SDSS J010013.02$+$280225.8 & $6.300\pm0.010$ & $-29.26\pm0.20$ & ... & \cite{Wu_etal:2015}\\
CFHQS J010250.64$-$021809.9 & $5.950\pm0.020$ & $-24.31\pm0.10$ & ... & \cite{Willott_etal:2009}\\
VHS J010953.13$-$304726.3 & $6.745\pm0.010$ & $-25.52\pm0.15$ & ... & \cite{Venemans_etal:2013}\\
CFHQS J013603.17$+$022605.7 & $6.210\pm0.020$ & $-24.40\pm0.10$ & ... & \cite{Willott_etal:2010a}\\
ATLAS J014243.70$-$332745.7 & $6.020\pm0.030$ & $-27.80\pm0.20$ & ... & \cite{Carnall_etal:2015}\\
ULAS J014837.64$+$060020.0 & $5.923\pm0.003$ & $-27.08\pm0.06$ & ... & \cite{Warren_etal:2015,Banados_etal:2014,Becker_etal:2015,Jiang_etal:2015}\\
ATLAS J015957.96$-$363356.9 & $6.310\pm0.030$ & $-27.00\pm0.10$ & ... & \cite{Carnall_etal:2015}\\
CFHQS J021013.19$-$045620.9 & $6.432\pm0.001$ & $-24.28\pm0.10$ & $6.41\pm0.02$ & \cite{Willott_etal:2010b,Willott_etal:2013}\\
CFHQS J021627.81$-$045534.1 & $6.010\pm0.020$ & $-22.21\pm0.10$ & ... & \cite{Willott_etal:2009}\\
CFHQS J022122.71$-$080251.5 & $6.161\pm0.014$ & $-24.45\pm0.10$ & ... & \cite{Willott_etal:2010a,Willott_etal:2010b}\\
PSO J022601.88$+$030259.4 & $6.541\pm0.002$ & $-27.36\pm0.03$ & $6.46\pm0.01$ & \cite{Venemans_etal:2015,Banados_etal:2014}\\
CFHQS J022743.29$-$060530.2 & $6.200\pm0.020$ & $-25.03\pm0.10$ & ... & \cite{Willott_etal:2009}\\
PSO J023152.96$-$285020.1 & $5.990\pm0.020$ & $-25.60\pm0.10$ & ... & \cite{Banados_etal:2014}\\
SDSS J023930.24$-$004505.4 & $5.820\pm0.030$ & $-24.50\pm0.12$ & ... & \cite{Jiang_etal:2009}\\
SDSS J030331.40$-$001912.9 & $6.070\pm0.001$ & $-25.82\pm0.10$ & $6.00\pm0.01$ & \cite{Jiang_etal:2008,Carilli_etal:2010}\\
VHS J030516.92$-$315056.0 & $6.604\pm0.008$ & $-25.96\pm0.06$ & ... & \cite{Venemans_etal:2013}\\
CFHQS J031649.87$-$134032.3 & $5.990\pm0.020$ & $-24.63\pm0.10$ & ... & \cite{Willott_etal:2010a}\\
VIKINGKIDS J0328$-$3252 & $5.850\pm0.030$ & ... & ... & \cite{Venemans_etal:2015b,Carnall_etal:2015}\\
VISTA J032835.51$-$325322.9 & $5.860\pm0.030$ & $-26.60\pm0.04$ & ... & \cite{Venemans_etal:2015b}\\
SDSS J035349.72$+$010404.4 & $6.049\pm0.004$ & $-26.49\pm0.08$ & $6.02\pm0.01$ & \cite{Jiang_etal:2008,Carilli_etal:2010}\\
DES J045401.79$-$444831.1 & $6.100\pm0.030$ & $-26.48\pm0.10$ & ... & \cite{Reed_etal:2015}\\
SDSS J081827.40$+$172251.8 & $6.020\pm0.020$ & $-27.40\pm0.10$ & $5.89\pm0.02$ & \cite{Fan_etal:2006c,Carilli_etal:2010}\\
ULAS J082813.42$+$263355.6 & $6.100\pm0.020$ & ... & ... & \cite{Warren_etal:2015}\\
SDSS J083643.86$+$005453.2 & $5.810\pm0.007$ & $-27.80\pm0.10$ & $5.62\pm0.01$ & \cite{Fan_etal:2001,Carilli_etal:2010}\\
VISTA J083955.3$+$001554.2 & $5.840\pm0.040$ & $-25.36\pm0.11$ & ... & \cite{Venemans_etal:2015b}\\
SDSS J084035.09$+$562419.9 & $5.844\pm0.002$ & $-26.60\pm0.10$ & $5.69\pm0.01$ & \cite{Fan_etal:2006c,Carilli_etal:2010}\\
SDSS J084119.52$+$290504.4 & $5.980\pm0.020$ & $-26.86\pm0.10$ & $5.81\pm0.01$ & \cite{Goto:2006,Carilli_etal:2010}\\
SDSS J084229.23$+$121848.2 & $6.080\pm0.020$ & $-27.18\pm0.03$ & ... & \cite{Fan_etal:2015,Wang_etal:2008}\\
SDSS J084229.43$+$121850.5 & $6.055\pm0.003$ & $-26.85\pm0.09$ & ... & \cite{Jiang_etal:2015}\\
SDSS J085048.25$+$324647.9 & $5.867\pm0.007$ & $-26.74\pm0.08$ & ... & \cite{Jiang_etal:2015}\\
SDSS J103027.10$+$052455.0 & $6.308\pm0.007$ & $-27.20\pm0.10$ & $6.21\pm0.01$ & \cite{Fan_etal:2001,Carilli_etal:2010}\\
SDSS J104845.05$+$463718.3 & $6.228\pm0.002$ & $-27.60\pm0.10$ & $6.16\pm0.01$ & \cite{Fan_etal:2003,Carilli_etal:2010}\\
CFHQS J104928.61$-$090620.4 & $5.920\pm0.020$ & $-25.58\pm0.10$ & ... & \cite{Willott_etal:2010a}\\
PSO J111033.98$-$132945.6 & $6.508\pm0.001$ & $-25.58\pm0.13$ & $6.48\pm0.01$ & \cite{Venemans_etal:2015}\\
ULAS J112001.48$+$064124.3 & $7.084\pm0.000$ & $-26.60\pm0.10$ & $7.04\pm0.01$ & \cite{Mortlock_etal:2011d,Venemans_etal:2012a}\\
SDSS J113717.70$+$354957.0 & $6.030\pm0.020$ & $-27.10\pm0.10$ & $5.91\pm0.01$ & \cite{Fan_etal:2006c,Carilli_etal:2010}\\
ULAS J114803.29$+$070208.3 & $6.200\pm0.020$ & ... & ... & \cite{Warren_etal:2015,Banados_etal:2014}\\
SDSS J114816.64$+$525150.3 & $6.419\pm0.002$ & $-27.80\pm0.10$ & $6.33\pm0.01$ & \cite{Fan_etal:2003,Carilli_etal:2010}\\
VISTA J114833.18$+$005642.2 & $5.840\pm0.030$ & $-24.46\pm0.11$ & ... & \cite{Venemans_etal:2015b}\\
ULAS J120737.44$+$063010.4 & $6.040\pm0.003$ & $-26.60\pm0.11$ & ... & \cite{Warren_etal:2015,Banados_etal:2014,Jiang_etal:2015}\\
PSO J121311.81$-$124603.5 & $5.860\pm0.020$ & $-27.30\pm0.10$ & ... & \cite{Banados_etal:2014}\\
VISTA J121516.87$+$002324.7 & $5.930\pm0.030$ & $-24.67\pm0.14$ & ... & \cite{Venemans_etal:2015b}\\
PSO J122913.21$+$041927.7 & $5.890\pm0.020$ & $-25.40\pm0.10$ & ... & \cite{Banados_etal:2014}\\
ULAS J124340.81$+$252923.9 & $5.830\pm0.020$ & ... & ... & \cite{Warren_etal:2015,Banados_etal:2014}\\
SDSS J125051.90$+$313022.0 & $6.150\pm0.020$ & $-27.10\pm0.10$ & $6.03\pm0.01$ & \cite{Fan_etal:2006c,Carilli_etal:2010}\\
SDSS J125757.47$+$634937.2 & $6.020\pm0.030$ & $-26.14\pm0.12$ & ... & \cite{Jiang_etal:2015}\\
SDSS J130608.26$+$035626.3 & $6.016\pm0.007$ & $-26.90\pm0.10$ & $5.92\pm0.01$ & \cite{Fan_etal:2001,Carilli_etal:2010}\\
ULAS J131911.29$+$095051.4 & $6.133\pm0.001$ & $-27.33\pm0.10$ & $6.04\pm0.01$ & \cite{Mortlock_etal:2009a,Wang_etal:2013,Banados_etal:2014}\\
SDSS J133550.81$+$353315.8 & $5.901\pm0.002$ & $-26.80\pm0.10$ & $5.89\pm0.01$ & \cite{Fan_etal:2006c,Carilli_etal:2010}\\
SDSS J140319.13$+$090250.9 & $5.860\pm0.030$ & $-26.27\pm0.11$ & ... & \cite{Jiang_etal:2015}\\
PSO J140329.33$-$120034.1 & $5.840\pm0.020$ & $-25.70\pm0.10$ & ... & \cite{Banados_etal:2014}\\
SDSS J141111.28$+$121737.3 & $5.904\pm0.007$ & $-26.80\pm0.10$ & $5.82\pm0.01$ & \cite{Fan_etal:2004,Carilli_etal:2010}\\
PSO J141327.12$-$223342.3 & $5.880\pm0.020$ & $-26.60\pm0.10$ & ... & \cite{Banados_etal:2014}\\
NDWFS J142516.30$+$325409.0 & $5.892\pm0.002$ & $-26.03\pm0.10$ & $5.76\pm0.01$ & \cite{Cool_etal:2006,Carilli_etal:2010}\\
FIRST J142738.59$+$331242.0 & $6.120\pm0.020$ & $-26.40\pm0.10$ & ... & \cite{McGreer_etal:2006}\\
CFHQS J142952.17$+$544717.7 & $6.210\pm0.020$ & $-25.85\pm0.10$ & ... & \cite{Willott_etal:2010a}\\
SDSS J143611.70$+$500707.0 & $5.850\pm0.020$ & $-26.30\pm0.10$ & $5.72\pm0.01$ & \cite{Fan_etal:2006c,Carilli_etal:2010}\\
CFHQS J150941.78$-$174926.8 & $6.121\pm0.003$ & $-26.98\pm0.10$ & ... & \cite{Willott_etal:2007,Willott_etal:2010b}\\
SDSS J160254.18$+$422822.9 & $6.090\pm0.020$ & $-26.80\pm0.10$ & $5.94\pm0.01$ & \cite{Fan_etal:2004,Carilli_etal:2010}\\
ELIAS J160349.07$+$551032.3 & $6.041\pm0.020$ & $-22.58\pm0.13$ & ... & \cite{Kashikawa_etal:2015}\\
ULAS J160937.28$+$304147.7 & $6.080\pm0.020$ & ... & ... & \cite{Warren_etal:2015}\\
SDSS J162331.81$+$311200.5 & $6.247\pm0.007$ & $-26.60\pm0.10$ & $6.16\pm0.01$ & \cite{Fan_etal:2004,Carilli_etal:2010}\\
SDSS J163033.90$+$401209.6 & $6.065\pm0.007$ & $-26.10\pm0.10$ & $5.94\pm0.01$ & \cite{Fan_etal:2003,Carilli_etal:2010}\\
CFHQS J164121.64$+$375520.5 & $6.047\pm0.003$ & $-25.48\pm0.10$ & ... & \cite{Willott_etal:2007,Willott_etal:2010a}\\
SDSS J205321.77$+$004706.8 & $5.920\pm0.030$ & $-25.47\pm0.07$ & ... & \cite{Jiang_etal:2009}\\
SDSS J205406.49$-$000514.8 & $6.039\pm0.000$ & $-26.11\pm0.09$ & $5.97\pm0.01$ & \cite{Jiang_etal:2008,Carilli_etal:2010,Wang_etal:2013}\\
CFHQS J210054.62$-$171522.5 & $6.087\pm0.003$ & $-25.03\pm0.10$ & ... & \cite{Willott_etal:2010a,Willott_etal:2010b}\\
SDSS J214755.41$+$010755.3 & $5.810\pm0.030$ & $-25.00\pm0.10$ & ... & \cite{Jiang_etal:2009}\\
VIMOS J221917.22$+$010248.9 & $6.156\pm0.020$ & $-23.10\pm0.11$ & ... & \cite{Kashikawa_etal:2015}\\
SDSS J222843.54$+$011032.2 & $5.950\pm0.020$ & $-24.50\pm0.10$ & ... & \cite{Zeimann_etal:2011}\\
CFHQS J222901.65$+$145709.0 & $6.152\pm0.003$ & $-24.52\pm0.10$ & ... & \cite{Willott_etal:2010a,Willott_etal:2010b}\\
PSO J223255.15$-$293032.2 & $6.658\pm0.007$ & $-26.04\pm0.09$ & $6.55\pm0.01$ & \cite{Venemans_etal:2015}\\
PSO J224048.98$-$183943.8 & $6.000\pm0.020$ & $-26.00\pm0.10$ & ... & \cite{Banados_etal:2014,Banados_etal:2014}\\
CFHQS J224237.55$+$033421.6 & $5.880\pm0.020$ & $-24.22\pm0.10$ & ... & \cite{Willott_etal:2010a}\\
SDSS J230735.35$+$003149.4 & $5.870\pm0.030$ & $-24.93\pm0.10$ & ... & \cite{Jiang_etal:2009}\\
SDSS J231038.89$+$185519.9 & $6.003\pm0.000$ & $-27.44\pm0.10$ & ... & \cite{Fan_etal:2015,Wang_etal:2013,Banados_etal:2014}\\
SDSS J231546.57$-$002358.1 & $6.117\pm0.006$ & $-25.38\pm0.08$ & $6.05\pm0.01$ & \cite{Jiang_etal:2008,Carilli_etal:2010}\\
CFHQS J231802.80$-$024634.0 & $6.200\pm0.020$ & $-24.83\pm0.10$ & ... & \cite{Willott_etal:2009}\\
CFHQS J232908.28$-$030158.8 & $6.417\pm0.002$ & $-25.23\pm0.10$ & $6.35\pm0.01$ & \cite{Willott_etal:2007,Willott_etal:2010a,Carilli_etal:2010}\\
CFHQS J232914.46$-$040324.1 & $5.900\pm0.020$ & $-24.36\pm0.10$ & ... & \cite{Willott_etal:2009}\\
VHS J234833.34$-$305410.0 & $6.886\pm0.009$ & $-25.72\pm0.14$ & ... & \cite{Venemans_etal:2013}\\
SDSS J235651.58$+$002333.3 & $6.000\pm0.030$ & $-24.92\pm0.10$ & ... & \cite{Jiang_etal:2009}\\
\end{longtable}
}

\section{Line-of-sight absorption by neutral hydrogren}
\label{section:los}

The most direct way in which quasars reveal
the (re-)ionization history of the Universe is as
sources of photons from which the absorption due to any
intervening neutral hydrogen can be inferred.
Critically, quasars have sufficiently similar spectra --
broad emission lines superimposed on a blue continuum -- that
the fraction of flux transmitted can be estimated reliably.
It is also important that the cosmological density of hydrogen
is sufficient to produce appreciable absorption even if the neutral
fraction is low;
conversely, there are some wavelengths at which only partial absorption
is produced even by a completely neutral IGM.


\subsection{Scattering of photons by a neutral hydrogen atom}
\label{section:lyman_series}

\begin{figure}[t]
\begin{center}
\includegraphics[scale=.6]{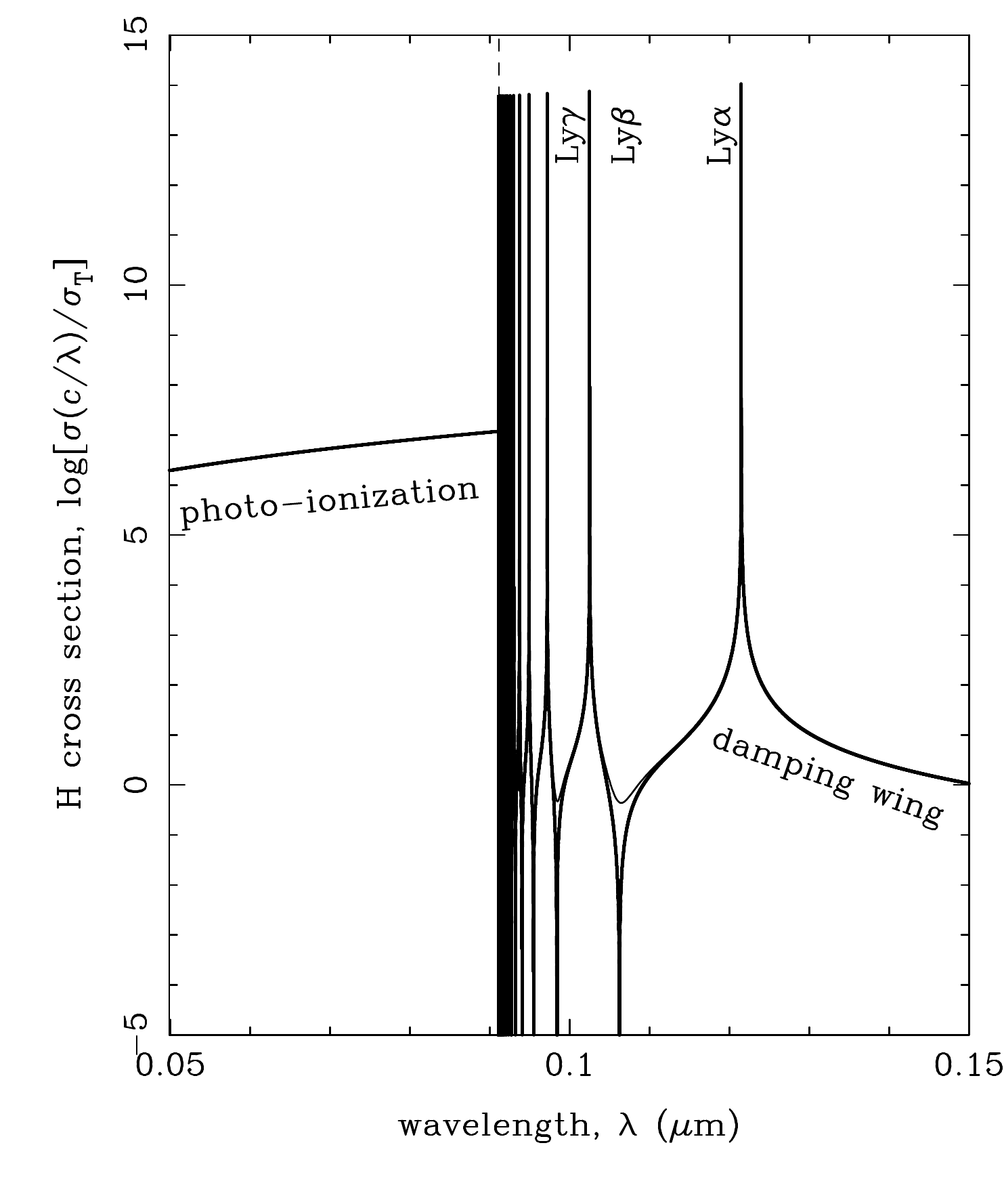}
\caption{Neutral hydrogen absorption cross section
  as a function of photon wavelength, $\lambda$.
  The Rayleigh scattering contribution
  (\eq{crosssection_full}, thick curve) dominates,
  although the additional contributions from Raman scattering can be
  seen between the \lya, \lyb\ and \lyg\ resonances (thin curve).
  (Despite appearances, the integrated cross section in a thin
  band of width $\Delta\lambda$ in the Lyman limit region near the 
  ionization threshold of $\lambda = 0.091 \unit{\micron}$ 
  is comparable to the smooth photo-ionization cross section.)}
\label{figure:crosssection}
\end{center}
\end{figure}

\begin{figure}[t]
\begin{center}
\includegraphics[scale=.6]{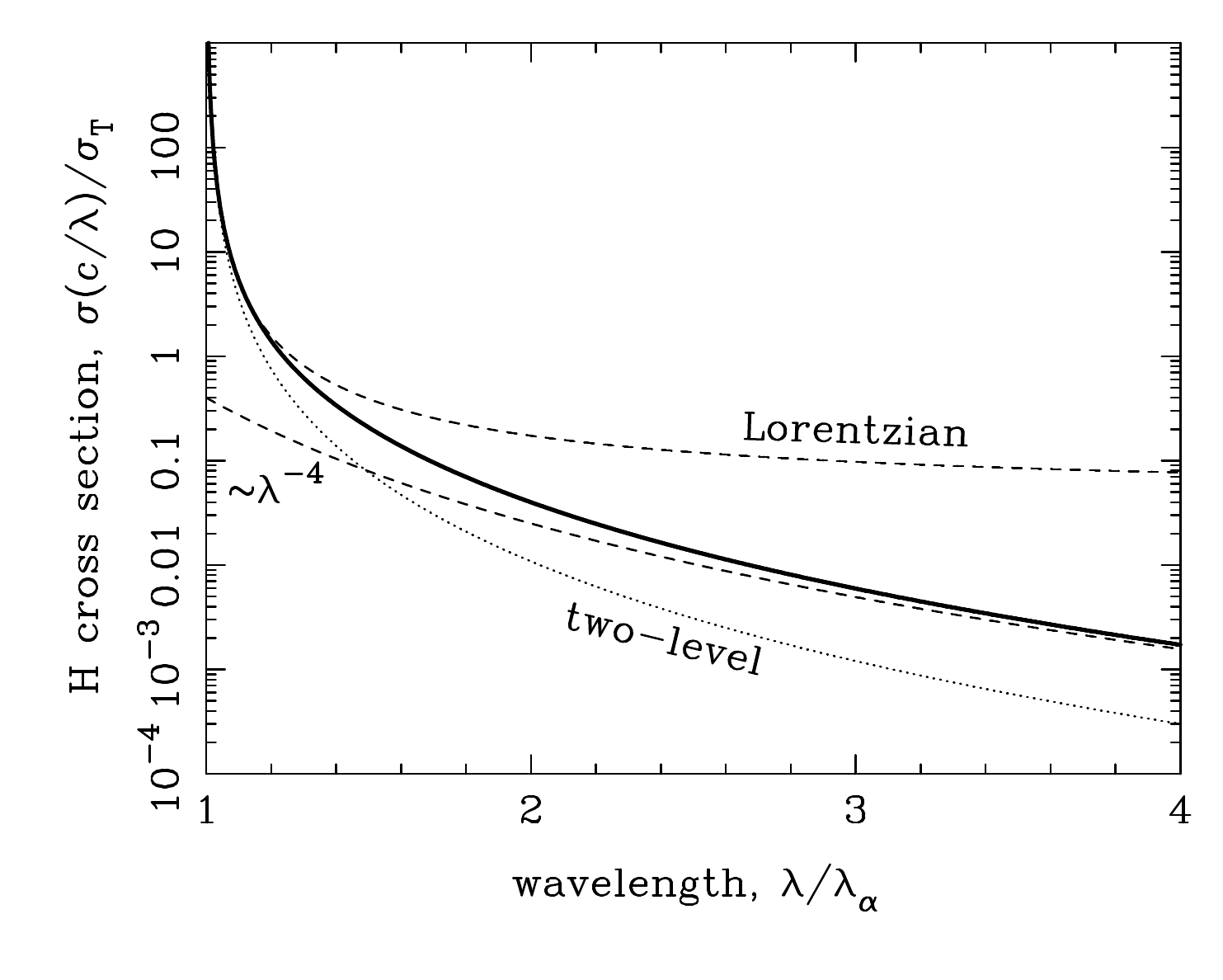}
\caption{\lya\ damping wing cross section, 
  shown as a function of photon wavelength, $\lambda$.
  The full cross section (\eq{crosssection_full}, solid curve)
  is compared to 
  the Lorentzian approximation (\eq{crosssection_lorentzian_nonres}, 
  dashed curve)
  and the low-frequency limiting form (\eq{rayleigh}, dashed curve)
  and 
  the two-level model (\eq{crosssection_peebles}, dotted curve) 
  commonly used in calculations of the IGM damping wing.}
\label{figure:crosssection_wing}
\end{center}
\end{figure}

A neutral hydrogen atom with its electron in the ground ($1s$)
state can absorb 
(and often re-emit, \ie, effectively scatter)
an incident photon
via a number of distinct quantum mechanical channels.
The resultant cross section has the form
\begin{equation}
\sigma(\nu) 
  = \sigma_{\rm Rayleigh}(\nu)
  + \sigma_{\rm Raman}(\nu)
  + \sigma_{\rm phot-ion}(\nu)
  + \ldots,
\end{equation}
where 
$\nu$ is the frequency\footnote{It would be 
more in keeping with astronomical conventions 
to give the cross section in terms of 
wavelength, $\lambda = c / \nu$;
and it is standard in quantum physics to 
use angular frequency, $\omega = 2 \pi \nu$,
or sometimes 
energy, $E = h \nu$. 
Frequency is used here it is is more directly linked to the 
physics of the scattering processes than wavelength,
while being more commonly used in astronomy than angular 
frequency.
The argument of the cross section
$\sigma(.)$ is {\em always} frequency here.}
of the incident photon
and the terms correspond to 
elastic Rayleigh scattering 
(in which the electron returns to the ground state)
inelastic Raman scattering
(in which the electron is left in an intermediate excited state),
photo-ionization
(in which the electron is removed from the atom), \etc\
The relative contributions of these processes are illustrated 
in \fig{crosssection}.
A first principles calculation
(\eg, \cite{Rybicki_DellAntonio:1994,Lee:2003,Hirata:2006})
requires evaluation of the
overlap integrals of the wave functions 
corresponding to the initial, intermediate (\ie, excited)
and final states,
but the essential phenomenology relevant to reionization 
studies can be understood without recourse to the full formalism.

\begin{table}
\caption{Properties of the Lyman series transitions}
\label{table:lyman_series}
\begin{tabular}{lrlrrr}
\hline\noalign{\smallskip}
name     & $n$      & transition          & wavelength, $\lambda_{1,n}$                  & decay rate, $\einstein_{1,n}$                 & oscillator strength, $f_{1,n}$ \\
\noalign{\smallskip}\svhline\noalign{\smallskip}
\lya     & 2        & $2p \rightarrow 1s$ & $0.12157 \unit{\micron}$                     & $6.265\times10^{8} \unit{s}^{-1}$             & 0.4162 \\
\lyb     & 3        & $3p \rightarrow 1s$ & $0.10257 \unit{\micron}$                     & $1.672\times10^{8} \unit{s}^{-1}$             & 0.0791 \\
\lyg     & 4        & $4p \rightarrow 1s$ & $0.97254 \unit{\micron}$                     & $0.682\times10^{8} \unit{s}^{-1}$             & 0.0290 \\
\lyd     & 5        & $5p \rightarrow 1s$ & $0.94974 \unit{\micron}$                     & $0.344\times10^{8} \unit{s}^{-1}$             & 0.0139 \\
$\cdots$ \\
         & $\gg1$   & $np \rightarrow 1s$ & $\frac{0.91175}{1 - 1 / n^2} \unit{\micron}$ & $\frac{41.80\times10^{8}}{n^3} \unit{s}^{-1}$ & $\frac{1.563}{n^3}$ \\
$\cdots$ \\
Ly limit & $\infty$ &                     & $0.91175 \unit{\micron}$                     &                                               & \\
\noalign{\smallskip}\hline\noalign{\smallskip}
\end{tabular}
\end{table}

Most important here is the Rayleigh 
scattering process,
which includes all the Lyman series transitions,
in particular the
strong \lya\ resonance
(in which a hydrogen atom's electron is excited from the 
$1s$ ground state to the $2p$ energy level and then returns to the ground
state) 
and the
slightly weaker \lyb\ 
($1s \rightarrow 3p$),
and \lyg\
($1s \rightarrow 4p$)
transitions.
These higher order transitions are also Raman scattering resonances, but 
the relative contribution from these channels is only significant away from the
resonances, as can be seen from \fig{crosssection}.
The $n$th electron energy level is 
$E_n = \rydberg (1 - 1 / n^2)$,
where 
$\rydberg
= m_{\rm e} \, e^4 / (8 \epsilon_0^2 \, h^2) \simeq 13.6 \unit{eV}$ 
is the Rydberg energy,
$m_{\rm e} = 9.109 \times 10^{-31} \unit{kg}$ is the electron mass,
$e = 1.602 \times 10^{-19} \unit{C}$ is the electron charge, 
$\epsilon_0 = 8.854 \times 10^{-12} \unit{F} \unit{m}^{-1}$ 
is the permittivity of free space\footnote{SI units are used here; in
cgs units $\epsilon_0 = 1 / (4 \pi)$ is effectively dimensionless.},
and 
$h = 6.626 \times 10^{-34} \unit{m}^2 \unit{kg} \unit{s}^{-1}$ 
is Planck's constant.
The fiducial wavelength and frequency
of a photon associated with the 
$1s \rightarrow np$ transition are hence
$\lambda_{1,n} = h c / [\rydberg (1 - 1 / n^2)]$
and 
$\nu_{1,n} = [\rydberg (1 - 1 / n^2)] / h$, respectively.
The spontaneous decay rate
(\ie, the Einstein $A$ coefficient) of the transition is 
\begin{equation}
\einstein_{1,n} = 
  \frac{\pi \, e^{10} \, f_{1,n} \, m_{\rm e}}
    {96 \, c^3 \, \epsilon_0^5 \, h^6}
  \, \left(1 - \frac{1}{n^2}\right)^2,
\end{equation}
with associated oscillator strength 
\begin{equation}
f_{1,n}
  = \frac{256 \, n^5 (n - 1)^{2n-4}}{3 (n + 1)^{2n + 4}}.
\end{equation}
These quantities are tabulated for the first few Lyman series transitions
in \tabl{lyman_series}.

The full Rayleigh scattering cross section
is obtained by squaring the sum over the complex amplitudes
associated with the various excited states
(as interference is possible),
the result of which can be approximated as 
(\cf\ \cite{Lang:1974,Alipour_etal:2015,Bach_Lee:2015})
\begin{equation}
\label{equation:crosssection_full}
\sigma(\nu)
  =
  \sigma_{\rm T}
  \left|\,
  \sum_{n = 2}^\infty f_{1,n}
      \frac{\nu^2}{\nu_{1,n}^2 - \nu^2 
     + i \, \einstein_{1,n} \nu_{1,n} / (2 \pi)}
  +
  \int_{\rydberg}^\infty
\frac{\diff f}{\diff E}
  \frac{1}{E^2 / (h \nu)^2 - 1}
 \, \diff E
  \,
  \right|^2,
\end{equation}
where 
$\sigma_{\rm T} =  
8 \pi / 3 [e^2 / (4 \pi \, \epsilon_0 \, m_{\rm e} \, c^2)]^2
\simeq 6.65\times10^{-29} \unit{m}^2$
is the Thomson cross section,
and the 
spectrum of oscillator strengths
for the unbound states is \cite{Alipour_etal:2015,Draine:2011}
\begin{equation}
\label{equation:dfde}
\frac{\diff f}{\diff E}
  = \step(E - \rydberg)
    \frac{128 \exp\{-4 \arctan[(E / \rydberg - 1)^{1/2}]
     / (E / \rydberg - 1)^{1/2}\}}
    {3 \rydberg (E / \rydberg)^4 
     \{1 - \exp[-2 \pi / (E / \rydberg - 1)^{1/2}]\}},
\end{equation}
with $\step(.)$ the Heaviside step function.
The first term in \eq{crosssection_full}
corresponds to the bound excited states
of the Lyman series.
This sum is dominated by the low-$n$ terms
and can be truncated in practice,
or evaluated 
using a Taylor series expansion \cite{Alipour_etal:2015} 
or a fitting formula \cite{Bach_Lee:2015,Mortlock_Hirata:2015}.

The main features of the Rayleigh scattering cross section,
shown in \fig{crosssection},
are 
the resonant Lyman series peaks and the long-wavelength
damping wing.
While these both come from the same underlying physics, 
their observational manifestations are sufficiently distinct that 
it is useful to consider them separately.

\subsubsection{Resonant absorption}

Near any of the Lyman series resonances 
the cross section given in \eq{crosssection_full} 
is dominated by a single term in the sum.
For the $1s \rightarrow np$ resonance this leads to 
\begin{eqnarray}
\sigma_{{\rm Lor},n}(\nu) 
\label{equation:crosssection_lorentzian}
  & \simeq & 
  \sigma_{\rm T} \, f_{1,n}^2
  \frac{\nu^4}{(\nu^2 - \nu^2_{1,n})^2 
    + \einstein_{1,n}^2 \, \nu^2_{1,n} / (4 \pi^2)}
\\
\label{equation:crosssection_lorentzian_nonres}
  & \simeq &
  \sigma_{\rm T} \frac{f_{1,n}^2}{4}
  \frac{\nu_{1,n}^2}{(\nu - \nu_{1,n})^2 + \einstein_{1,n}^2 / (16 \pi^2)}
\mathif
  |\nu - \nu_{1,n}| \ll \nu_{1,n},
\end{eqnarray}
the Lorentzian form that is often used to analyse discrete
\hi\ concentrations (such as in the \lya\ ``forest'' 
and the high column density systems considered in \sect{wing}).
These resonances are extremely sharp --
for the \lya\ transition 
$\einstein_\alpha / \nu_\alpha \simeq 10^{-7}$,
so 
the peak at a wavelength of $\lambda_\alpha \simeq 0.12 \unit{\micron}$
has an characteristic width of just $\sim 10^{-8} \unit{\micron}$  --
and are completely unresolved in \fig{crosssection}.
The cross section at resonance is correspondingly high,
being given almost exactly by
$\sigma(\nu_{1,n}) 
  = 4 \pi^2 \, \sigma_{\rm T} \, f_{1,n}^2 \, \nu_{1,n}^2
  / \einstein_{1,n}^2$.
In the case of \lya\ this is 
$\sigma_\alpha \simeq 4.41 \times 10^{-16} \unit{m}^2$.

While there is no physical 
reason that the integral of $\sigma(\nu)$
over wavelength should be finite -- 
indeed,
the full form of \eq{crosssection_full}
cannot be normalised to give a probability distribution in $\nu$ -- 
adopting \eq{crosssection_lorentzian_nonres} implies that 
$\int_0^\infty \diff \nu \, \sigma_{{\rm Lor},n}(\nu) 
\simeq 
\int_{-\infty}^\infty \diff \nu \, \sigma_{{\rm Lor},n}(\nu)
= \pi^2 \, \sigma_{\rm T} \, f_{1,n}^2 \, \nu_{1,n}^2 
  / \einstein_{1,n}$.
Treating 
the absorption at each resonance as completely mono-chromatic 
then leads to the useful approximation that the full cross
section is
\begin{equation}
\label{equation:crosssection_delta}
\sigma(\nu) 
  \simeq 
  \sigma_{\rm T} 
    \sum_{n = 2}^\infty
    \frac{\pi^2 \, f_{1,n}^2 \, \nu^2_{1,n}}{\einstein_{1,n}} \,
    \diracdelta(\nu - \nu_{1,n}),
\end{equation}
where $\diracdelta(.)$ is the Dirac delta function.
This is used in the standard
derivation of the IGM optical depth given in \sect{gp}.


\subsubsection{Damped absorption}

Redward of the \lya\ resonance, the damped absorption wing
includes contributions from all the excited states.
The full damping wing cross section
(\ie, extending to the \lya\ resonance)
can be evaluated numerically
from the formula given in \eq{crosssection_full},
although
transmission is typically close to zero 
just redward of the \lya\ line
so there is little practical need to match the
in-resonance form of $\sigma(\nu)$ precisely
in calculations of the damping wing absorption (\sect{wing}).
This opens up the possibility of using 
one of several available approximations
\cite{Alipour_etal:2015,Bach_Lee:2015,Mortlock_Hirata:2015}.

In the low frequency limit 
the $\nu$-dependence of all the
terms in \eq{crosssection_full} is such that the classical
$\sigma(\nu) \propto \nu^4$ 
Rayleigh scattering result holds for each individually;
they sum to give \cite{Alipour_etal:2015,Bach_Lee:2015}
\begin{equation}
\label{equation:rayleigh}
\sigma(\nu) \simeq 0.399 \,
\sigma_{\rm T} 
\left(\frac{\nu}{\nu_\alpha}\right)^4
\mathif
\nu \ll \nu_\alpha,
\end{equation}
where $\nu_\alpha = \nu_{1,2}$ is the frequency of the \lya\
resonance.

That said, 
extrapolation of the $n = 2$ Lorentzian profile of 
\eq{crosssection_lorentzian_nonres} 
to lower frequencies, which gives
\begin{equation}
\sigma_{{\rm Lor},2} (\nu) \simeq 
 \sigma_{\rm T} \, f_{1,n}^2
  \frac{1}{4 \, (\nu /\nu_{\alpha} - 1)^2 }
\mathif
\nu - \nu_\alpha \ll \einstein_{1,2},
\end{equation}
does not yield a good approximation to the 
damping wing,
as can be seen from \fig{crosssection_wing}.
This point was
made by Miralda-Escud\'{e} \cite{Miralda-Escude:1998},
who instead advocated using the form of $\sigma(\nu)$
derived by Peebles \cite{Peebles:1993}
and based in turn on a classic quantum mechanics calculation
\cite{Weisskopf_Wigner:1930}.
The underlying model is that
a hydrogen atom is a two-level system
in which a bound electron is restricted to the 
$1s$ or $2p$ states, 
which yields a cross section of
\cite{Peebles:1993,Miralda-Escude:1998,Weisskopf_Wigner:1930}
\begin{eqnarray}
\sigma_{\rm 2L} (\nu) 
\label{equation:crosssection_peebles}
  & = & 
  \sigma_{\rm T} \frac{f_{1,2}^2}{4}
  \frac{\nu_\alpha^2 \, (\nu / \nu_\alpha )^4}
    {(\nu - \nu_\alpha )^2
    + \einstein_{1,2}^2 / (16 \pi^2) \, (\nu / \nu_\alpha )^6}
\mathif
\nu \ga \nu_\alpha
\\
\label{equation:crosssection_miralda}
& \simeq & 
  \sigma_{\rm T} \, f^2_{1,2}
  \frac{ (\nu / \nu_\alpha )^4}
    {4 \,(\nu /\nu_\alpha - 1 )^2}
\mathif
\nu - \nu_\alpha \ll \einstein_{1,2}.
\end{eqnarray}
The first formula includes the \lya\ resonance,
correctly 
matching 
\eq{crosssection_lorentzian_nonres} if $\nu \simeq \nu_\alpha$;
and the $(\nu / \nu_\alpha)^4$ term in 
the numerator of both expressions gives a full damping wing.
In the low frequency limit,
however,
\begin{equation}
\sigma_{\rm 2L} (\nu)
  \simeq 0.0433 \, \sigma_{\rm T} \left( \frac{\nu}{\nu_\alpha}\right)^4
\mathif
\nu \ll \nu_\alpha,
\end{equation}
which is a factor of $\sim 10$ smaller than the correct
limit given in \eq{rayleigh}.
The reason 
\cite{Peebles:1993,Bach_Lee:2015}
for this discrepancy 
is the omission of the contributions from the other excited
states ($3p$, $4p$, \etc, as well as the unbound continuum), 
all of which contribute to Rayleigh scattering.

Rather ironically, the Lorentzian form given in
\eq{crosssection_lorentzian_nonres} is, in practice,
a more useful approximation to the
red damping wing of the \lya\ line.
Even though the implied low frequency cross section 
has the wrong limiting form,
with $\sigma_{{\rm Lor},n}(\nu) \rightarrow 0.0433 \, \sigma_{\rm T}$
as $\nu \rightarrow 0$,
it is actually a good match for
$0.9 \, \nu_\alpha \la \nu \leq \nu_\alpha$
which dominates the absorption.
The implications of using different forms for $\sigma(\nu)$ to model the 
IGM damping wing are discussed in \sect{wing}.


\subsection{Absorption by neutral hydrogen in an expanding universe}
\label{section:expand}

The optical depth from the \hi\ between observer and source 
is given by integrating the absorption cross section
along the line-of-sight
and also integrating 
over the (line-of-sight) velocity distribution at each point,
although it is useful to consider the velocity integral as
a separate convolution.
A collection of \hi\ atoms with 
(normalised) line-of-sight velocity distribution 
$\phi(\vlos)$ can be treated as if {\em each} has the effective cross section
\begin{equation}
\label{equation:sigma_eff}
\sigma_{\rm eff} (\nu)
  \, = \, \int_{-c}^c
    \diff \vlos \, \phi(\vlos) \, 
  \sigma\!\left(\frac{\nu}{1 + \vlos / c}\right)
  \, \simeq \,
     \int_{-\infty}^\infty
    \diff \vlos \, \phi(\vlos) \, 
  \sigma\!\left[\nu (1 - \vlos / c)\right]
\end{equation}
where $\vlos$ is defined to be positive away from the observer
and it is assumed that the motions are non-relativistic.

If the motions are thermal then it is reasonable to adopt
a Gaussian velocity distribution, 
$\phi(\vlos) \, = \,
  \exp[ -\vlos^2 / (2 \sigmalos^2)] / [(2 \pi)^{1/2} \, \sigmalos]$,
where $\sigmalos \ll c$ is the temperature-dependent 
(line-of-sight) velocity dispersion.
The resultant form of $\sigma_{\rm eff}(\nu)$ 
differs significantly from $\sigma(\nu)$ 
only near the sharp Lyman series resonances.
The Lorentzian line profile of the $1s \rightarrow np$
transition given in 
\eq{crosssection_lorentzian_nonres}
hence becomes 
\begin{equation}
\label{equation:voigt}
\sigma_{{\rm eff},1,n}(\nu)
  \, = \, \sigma_{\rm T} \,
    \frac{\pi^2 \, f_{1,n}^2 \, \nu_{1,n}^2}{\einstein_{1,n}}
\end{equation}
\[
\mbox{} \times
   \int_{-\infty}^\infty \diff \nu^\prime
   \frac{1}{(2 \pi)^{1/2} \, \nu_{1,n} \, \sigmalos / c}
   \exp\left[
      -\frac{1}{2} \left(\frac{\nu - \nu^\prime}{\nu_{1,n} \, \sigmalos / c} 
   \right)^2\right]
  \frac{\einstein_{1,n} / (4 \pi)}
   {\pi [(\nu^\prime - \nu_{1,n})^2 - \einstein_{1,n}^2 / (4 \pi)^2]}
,
\]
the Voigt profile form that is valid if
$\sigma \ll c$ and 
$\einstein_{1,n} \ll \nu_{1,n}$
(conditions which imply that the cross section is only significant 
for $\nu \simeq \nu_{1,n}$).
The integral cannot be evaluated analytically, 
but a number of useful numerical approximation methods
have been developed 
(\eg, 
\cite{Davies_Vaughn:1963,Tepper-Garcia:2006,Zaghloul:2007,Schroeder_etal:2013}).
The second line of
\eq{voigt} is the convolution of a normal distribution
and a Cauchy distribution, and so is itself
a normalised Voigt distribution in $\nu$,
centred at $\nu_{1,n}$.
Far from resonance the Voigt profile has the same heavy tails
as the Lorentzian, but close to resonance the peak
is broadened to have Gaussian ``core'' of standard deviation
$\sigma_\nu \simeq \nu_{1,n} \, \sigmalos / c$.
In the context of quasar reionization studies this 
is important for modelling the absorption
by residual \hi\ in quasar \hii\ zones
(\sect{nearzone}) 
and also for correctly calculating the 
near-resonance form of the 
\lya\ damping wing (\sect{wing}).

For most of the results presented here,
however, line-of-sight \hi\ velocities can be ignored,
either because the detailed shape of the
resonant line profile itself is irrelevant
(\eg, \sect{gp}) or because it is only the 
long-wavelength limit of the smooth damping wing that
is important
(most of \sect{wing}).
In this case the simplest approach is to 
assume that $\phi(\vlos) = \diracdelta(\vlos)$,
in which case 
\begin{equation}
\sigma_{\rm eff}(\nu) = \sigma(\nu).
\end{equation}

The infintesimal optical depth 
to photons of wavelength $\lambda$
along a (proper)
line-of-sight interval of length $|\diff l|$
in a region with a local (\ie, proper) 
density $\nhilos$ of neutral hydrogen is
\begin{equation}
\label{equation:dtau_full}
\diff \tau (\lambda) = 
  \nhilos \, \sigma_{\rm eff} \left( \frac{c}{\lambda} \right) \, |\diff l|,
\end{equation}
where $\sigma_{\rm eff}(\nu)$ is defined in \eq{sigma_eff}.
Integrating \eq{dtau_full} along the line-of-sight
from source to observer
gives the optical depth 
as a function of 
observed wavelength $\lambda_\obs$
emitted
by a source at 
redshift $\redshift_\src$ 
as
\begin{eqnarray}
\label{equation:tau_general}
\tau (\lambda_\obs)
  & = & \int_0^{\redshift_\src} \diff \redshift \, 
    \frac{c}{(1 + \redshift) \, H(\redshift)}
    \nhi(\redshift)
    \,
    \sigma_{\rm eff}\!\left(\frac{1 + \redshift}{\lambda_\obs / c} \right)
    \\
  & \simeq & 
\label{equation:tau_highz}
   \frac{c \, \bar{n}_{\rm H,0}}{\Omega_{\rm m}^{1/2} \, H_0}
   \int_0^{\redshift_\src} \diff \redshift \, 
     (1 + \redshift)^{1/2} \,
      \fhi(\redshift) \,
     \sigma_{\rm eff}\!\left(\frac{1 + \redshift}{\lambda_\obs / c} \right),
\end{eqnarray}
where 
\eq{dl} has been used to change the integration variable
from $l$ to $\redshift$ 
and 
the second expression uses both the high-redshift
$H(\redshift)$ relationship
given in \eq{h_z}
and the cosmological form of the \hi\ density
given in \eq{nhicosmo}.

The fraction of light transmitted is then
\begin{equation}
\label{equation:transmit}
T(\lambda_\obs) = \exp[- \tau(\lambda_\obs)].
\end{equation}
\Eq{transmit}
is the basis for the majority of the
methods described below to use absorption measurements of quasars
to study the 
reionization history of the Universe.


\section{The Gunn-Peterson effect}
\label{section:gp}

\begin{figure}[t]
\begin{center}
\includegraphics[scale=.6]{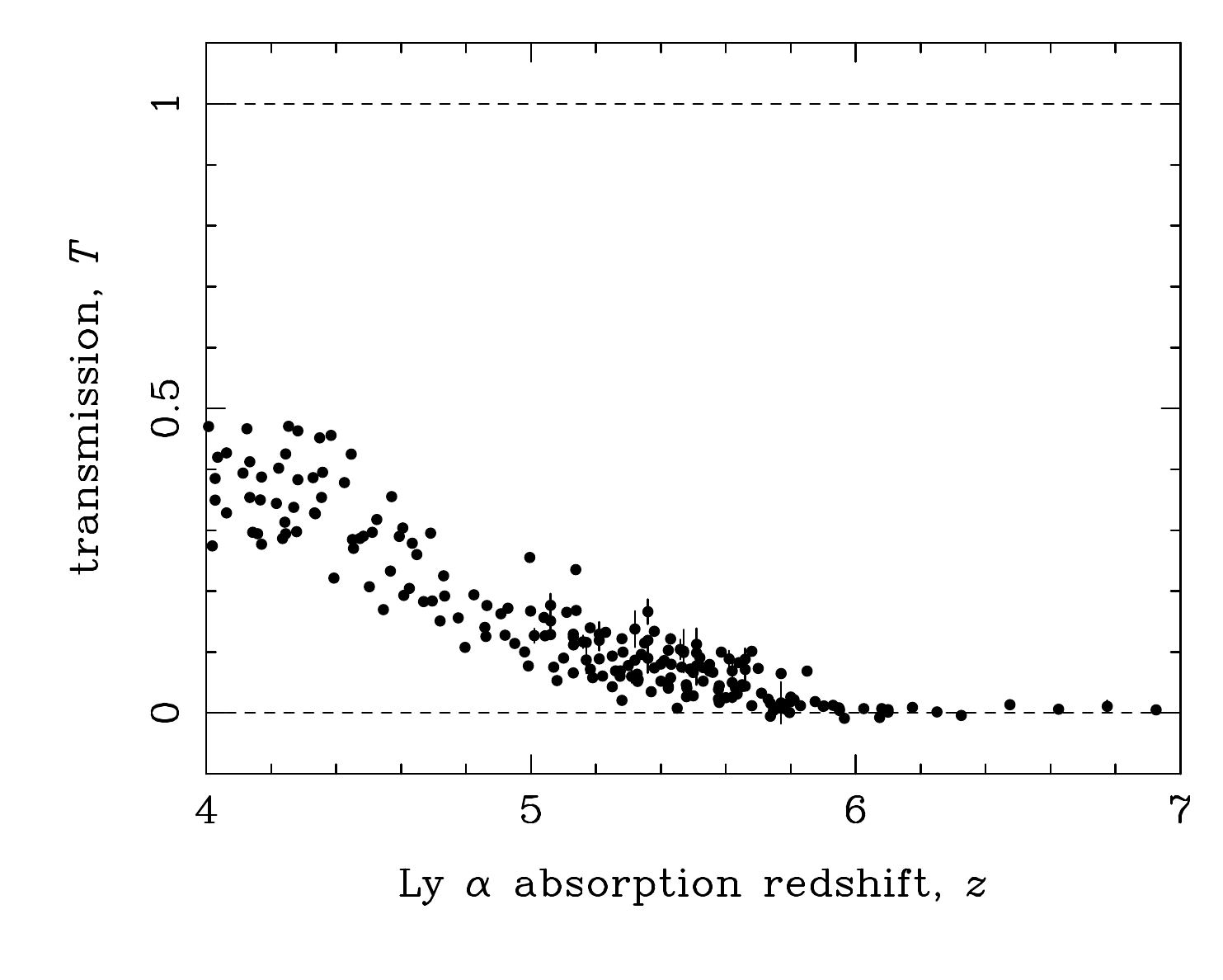}
\caption{Measured transmission, $\hat{T}$,
  along the lines of sight to \hzqs\ at a range of redshifts
  \cite{Fan_etal:2006b,Mortlock_etal:2011d,Becker_etal:2015,Songaila:2004},
  shown as a function of \lya\ redshift, 
  $\redshift = \lambda_\obs / \lambdaalpha - 1$.}
\label{figure:transmission}
\end{center}
\end{figure}

The Lyman series cross sections are so large 
near resonance that the IGM would, if even slightly neutral,
be all but opaque to all photons with wavelengths of 
$\lambda \leq \lambda_\alpha$.
This remarkable fact was discovered independently 
several times 
\cite{Field:1959,Shklovskii:1964,Scheuer:1965,Bahcall_Salpeter:1965},
but it was not until
Gunn \& Peterson \cite{Gunn_Peterson:1965} 
presented their calculation, along with 
measurements \cite{Schmidt:1965}
of a $\redshift = 2.01$ quasar,
that the significance of this result was broadly appreciated,
and it is now known almost universally as the Gunn-Peterson (\gp) effect.

The opacity of the IGM is 
independent of the detailed wavelength dependence of the 
Lyman series resonances (\sect{lyman_series}), 
and can be calculated 
by adopting the delta function approximation to 
$\sigma(\nu)$
given in \eq{crosssection_delta}.
The integral in \eq{tau_general}
can then be evaluated directly;
converting from observed wavelength to absorption redshift
according to $\redshift_n = \lambda_\obs / \lambda_{1,n} - 1$ 
gives the \gp\ optical depth associated with the 
$n$th Lyman series resonance as
\begin{equation}
\label{equation:tau_gp}
\tau_{\gp,n}(\redshift_n) 
  = 
    \frac{\pi^2 \, c^2 \, \sigma_{\rm T} \, f_{1,n}^2 \, \nhi(\redshift_n)}
    {\lambda_{1,n} \, \einstein_{1,n} \, H(\redshift_n)} .
\end{equation}
The resultant total optical depth to a source at redshift $\redshift$
is then obtained by the summing over the Lyman series to give
\begin{equation}
\tau(\lambda_\obs)
  = \sum_{n = 2}^{\infty}
    \step[(1 + \redshift) \, \lambda_{1,n} - \lambda_\obs]
    \,\,
    \tau_{\gp,n}\!\left(\frac{\lambda_\obs}{\lambda_{1,n}} - 1 \right),
\end{equation}
where the step functions produce a distinctive
``saw-tooth'' pattern in $\tau(\lambda_\obs)$ 
as the different resonances drop out with increasing wavelength
(\cf\ \cite{Madau:1995,Roche_etal:2012}).

Most important is the optical depth that results from the \lya\
resonance, both because it is the strongest of the Lyman series
transitions and has the longest wavelength 
(hence being by far the dominant channel by which photons
emitted with $\lambda_{\rm em} > \lambda_\beta \simeq 0.1 \unit{\micron}$
are absorbed).
Evaluating \eq{tau_gp} 
for $n = 2$ and adopting the fiducial values for the cosmological
parameters from \sect{cosmology} then gives
\begin{equation}
  \tau_{\gp,\alpha}(\redshift)
  \simeq 
  3.3 \times 10^5
  \frac{H_0}{70 \unit{km} \! \unit{s}^{-1} \! \unit{Mpc}^{-1}}
  \left(\frac{\Omega_{\rm m}}{0.3}\right)^{-1/2}
  \frac{\Omega_{\rm b}}{0.04}
  \left(\frac{1 + \redshift}{7}\right)^{3/2}
  \fhi(\redshift). 
\end{equation}
Thus 
a very low IGM neutral fraction of $\fhi \ga 10^{-5}$ would be sufficient
to result in significant absorption;
and if $\fhi \ga 10^{-3}$ there would be 
almost complete absorption of all photons 
emitted with $\lambda < \lambda_\alpha$.
The \gp\ effect is hence best suited to probing the end of
reionization, as even deep spectroscopic observations of the
brightest sources would only result in lower limits on
$\tau_\gp$ and hence $\fhi$ (\cf\ \cite{Mortlock_etal:2011d,Tanvir_etal:2009}).
As conceptually simple as this probe of the IGM is, 
there are a number of complications that make the interpretation
of such continuum transmission measurements somewhat unclear.

One source of ambiguity is in simply estimating the 
fraction of the flux transmitted,
naturally estimated as 
\begin{equation}
\label{equation:transmission}
\hat{T}_n(z) = \frac{\hat{F}_\nu[(1 + \redshift) \, \lambda_{1,n}]}
  {F_{{\rm intr},\nu}[(1 + \redshift) \lambda_{1,n}]},
\end{equation}
where $\hat{F}_\nu(\lambda)$ is the measured flux density of the 
quasar and $F_{{\rm intr},\nu}(\lambda)$ is a model 
of the quasar's intrinsic spectrum.
This is usually assumed to be a power-law
(\eg, \cite{Fan_etal:2006b}), 
and while the results are somewhat sensitive to the 
model adopted \cite{Kim_etal:2007,Lee:2012},
the increase of $\tau_\gp$ with $\redshift$ is sufficiently
dramatic that the uncertainties in the intrinsic emission do
not affect the broad interpretation of the transmission estimates.

A second issue is the possibility that the measured absorption 
is not entirely due to \hi\ in the IGM.  Specifically, line blanketing
\cite{Madau:1995} has been invoked 
as an explanation for the high absorption towards some \hzqs.

Regardless of the cause of the absorption, it has become standard 
to define the observable effective GP optical depth in terms 
of the estimated transmission (defined in \eq{transmission}) as 
\begin{equation}
\hat{\tau}_{\rm GP,eff}(\redshift) = -\ln[\hat{T}(\redshift)] .
\end{equation}
In the high-$\tau$ limit this quantity has awkward statistical 
properties, even being undefined if $\hat{T}(\redshift) \leq 0$,
but it is a good summary provided it is evaluated over a sufficiently
broad spectral range that non-zero flux is detected.
\Fig{transmission} illustrates the observed evolution 
in the observed transmission with redshift,
and also that there is significant variation between
different lines-of-sight.

At redshifts of $\redshift \la 5$ the dominant \lya\
absorption is in the form discrete \hi\ concentrations
that are associated with collapsed objects
and seen as the \lya\ forest.  
The level of the continuum between these absorption lines 
can then be used to estimate -- or at least place limits on --
the GP optical depth, 
although the distinction between continuum/IGM 
and halos/clouds is somewhat artificial 
(\eg, \cite{Cen_etal:1994}).
Still, this technique has usefully been applied to increasingly 
distant quasars, 
yielding, \eg,
$\tau_{\rm GP,eff} \la 0.05$ at $\redshift \simeq 4$ 
\cite{Webb_etal:1992,Giallongo_etal:1994},
$\tau_{\rm GP,eff} \simeq 0.1$ at $\redshift = 4.8$ 
\cite{Songaila_etal:1999}.
and 
$\tau_{\rm GP,eff} \simeq 0.4$ at $\redshift = 5.5$ 
\cite{Fan_etal:2000}.
At higher redshifts the \lya\ forest lines overlap to the 
degree that identifying regions of continuum emission becomes 
impractical.

The most comprehensive source of information about the IGM 
optical depth at $\redshift \simeq 6$ comes from the 
19 bright $5.7 \la \redshift \la 6.5$ analysed by
Fan \etal\ \cite{Fan_etal:2006b}. 
The marked increase in $\hat{\tau}_{\rm GP,eff}$ 
with $\redshift$ beyond that which would be expected from
the increase in proper density is generally taken as 
a clear identification of the tail-end of the cosmological
reionization, although this conclusion does rest on 
a number of assumptions 
about the distribution of \hi\ in the IGM and the ionizing background
(\cf\ \cite{Mesinger_Haiman:2007,Mesinger:2010}).

At redshifts of $\redshift \ga 6$,
the density of \hi\ was sufficiently high 
that $\tau_\gp \gg 1$, 
meaning that almost all light
blueward of \lya\ is absorbed.
The GP effect cannot be exploited to 
measure $\fhi$ deep into the reionization epoch, 
as it is only possible to place lower limits on $\tau_\gp$ 
and $\fhi$.  
To probe further -- at least using quasars -- 
requires going beyond 
the average properties of the IGM
(as encoded by the optical depth) and exploiting its
structure (\sect{gaps}),
the distribution of transmission values (\sect{darkpixel}),
or 
the comparatively weak red damping wing of the \lya\ 
transition (\sect{wing}).  


\subsection{Dark gaps}
\label{section:gaps}

If, at a given redshift,
the neutral fraction of the hydrogen in the IGM 
was $\fhi \ga 0.01$ then almost all photons 
emitted with $\lambda < \lambda_\alpha$
from sources at or beyond this redshift would absorbed.
As discussed in \sect{gp}, the resultant GP 
optical depth measurements
cannot be used to do anything more than place upper limits on $\fhi$.  
One alternative 
(\cite{Croft:1998,Barkana:2002,Songaila_Cowie:2002}), 
is to instead look at the lengths of the GP 
troughs, by applying what has become known as a dark gap analysis.

Assuming the reionization process is patchy, 
it is expected that even at times when the Universe was
still fairly neutral there would be small highly-ionized regions 
(\eg, \cite{Iliev_etal:2006}),
which might be seen as sharp ``spikes'' of transmission in the 
spectra of \hzqs.
(Early in the reionization process it is possible that the 
even highly ionized regions have a sufficient residual density of 
\hi\ to be opaque to \lya\ photons, 
and the spatial fluctuations in the ionizing background 
\cite{Becker_etal:2015} further complicate this issue.)
There is expected to be more information in the positions of
these spikes -- and, in particular, the separations between them --
than can be encoded by simply folding them into an optical 
depth measurement.
Between these transmission peaks would be long regions with no 
detectable flux; these dark gaps are expected to be longer
at higher redshifts when the IGM was more neutral and 
so could 
are the extreme version of the GP troughs discussed in 
\sect{gp}, and are often referred to as dark gaps.
The potential importance of dark gaps lies in the possibility
that their number and length might be both observable and 
sensitive probes of the ionization state of the IGM;
to date, however, it has proved difficult to convert this 
appealing idea into quantitive constraints on the cosmological
reionization history.

One difficulty in all dark gap studies is that the definition is,
unlike that of, \eg, the GP optical depth,
dependent on the observation under consideration.
The identification of a transition spike,
which would hence break a long dark gap into two shorter gaps, 
does not necessarily correspond to real transmission, 
and so this technique cannot be 
completely disentangled from the noise level and 
resolution of the spectrum
being analysed.
Particularly in the case of $\redshift \ga 7$
sources, the residual presence of sky lines in the relevant 
wavelength range (\eg, \cite{Mortlock_etal:2011d})
means dark gap statistics are subject to a very complicated
source of noise.
In more concrete terms, two different observations of the same
quasar could easily yield very different dark gap numbers and lengths,
implying that these statistics are not robust.

Observations of redshift $\la 5$ quasars
showed increasing absorption due to isolated concentrations
of \hi, but the first detection \cite{Djorgovski_etal:2001}
of an extended region of high absorption was in the 
$\redshift = 5.74$ quasar SDSS~J1104$-$0125, 
for which $\tau_{\rm GP,eff} > 4.6$ 
over the redshift range $5.2 \la \redshift \la 5.6$.
This was not, however, unequivocal evidence of an increase
in the IGM neutral fraction as line blanketing 
(\cf\ \sect{gp}) could also result in long sections of 
minimal transmission from a region of IGM that was mostly,
by volume, ionized \cite{Barkana:2002}.
Subsequent observations 
of the 
$\redshift = 6.30$ quasar SDSS~J1030+0524 \cite{Becker_etal:2001}
and the 
$\redshift = 6.42$ quasar SDSS~J1148+5251 \cite{Fan_etal:2003}
revealed even longer dark gaps, 
spanning the redshift ranges
$6.0 \leq \redshift \la 6.26$ \cite{White_etal:2003,Becker_etal:2001}
and 
$6.1 \leq \redshift \la 6.30$ \cite{White_etal:2003},
respectively.
The expected increase in gap length with redshift
was confirmed by the 
$\redshift = 7.08$ quasar ULAS~J1120+0641 \cite{Mortlock_etal:2011d},
which shows no detectable \lya\ transmission 
above $\redshift \simeq 5.8$, albeit from a noisier
spectrum than those available for 
SDSS~J1030+0524 
and 
SDSS~J1148+5251.
(There is also no emission detected blueward of the \lya\ lines of the
three $\redshift > 6.5$ quasars discovered in VISTA \cite{Venemans_etal:2013},
although the spectra published to date are noisier again.)

While there is a clear trend with redshift, there is 
also signficant variation between lines-of-sight 
(\eg, \cite{Fan_etal:2006b}), as shown most strikingly by the 
discovery \cite{Becker_etal:2015} of a $\tau_{\rm GP,eff} > 7$ trough
covering the redshift range $5.5 \la \redshift \la 5.9$ in the 
$\redshift = 5.98$ quasar ULAS~J0148+0600 \cite{Warren_etal:2015}.
The strongest inference that can be made from these observations
is that reionization was an extremely inhomogeneous
process, counselling against making any strong inferences 
from measurements along a single line-of-sight.

Moving from the qualitative result that the dark gap lengths 
increase with redshift to quantitative constraints on the 
neutral fraction has also proved diffuclt. 
In principle, the best option would be to compare the 
gap distribution with the predictions of numerical
models 
\cite{Paschos_Norman:2005,Kohler_etal:2005,Gallerani_etal:2006},
but great care must be taken to avoid making strong conclusions
that are tied to assumptions about the spatial distribution of 
hydrogen and the ionizing background.
Partly for this reason,
the promised quantitative constraints 
on the neutral fraction have not been forthcoming;
but dark gaps might become more important
as spectra are obtained of quasars deeper into the reionization 
epoch.


\subsection{Dark pixels}
\label{section:darkpixel}

Another way to exploit the almost complete 
absorption produced by regions of \hi\
is to look at the fraction of of a \hzq\
spectrum for which there is no (detectable) transmission.
The key point here is that the transmission distribution is,
by virtue of its exponential relationship to the \hi\ density,
strongly skewed to values close to zero; with the addition 
of observational noise there is a 50\% chance that any such pixels 
will have negative measured transmission.

It has been argued \cite{McGreer_etal:2011,Mesinger:2010}
that dark gap statistics provide the most model-independent
constraints on the neutral fraction,
especially towards the end of reionization.
The inevitable price is a loss in precision,
a result of discarding a large amount of the information
contained in the observed transmission profiles,
but in such a complicated setting reliability is to be valued.
The most extreme version of this trade-off is to estimate 
the number of dark pixels as twice the number 
of pixels with negative measured fluxes
(\eg, \cite{McGreer_etal:2011}).
Applying this methodology to 22 redshift $\sim 6$
quasar spectra gave
$\fhi \leq 0.06 \pm 0.05$ 
at a redshift of 5.9,
robust additional 
evidence that reionization was close to complete
by $\redshift \simeq 6$.
\cite{McGreer_etal:2015}.


\section{Quasar H {\small \bf II}  regions}
\label{section:nearzone}

Quasars are not only passive probes of the \hi\ density, 
but also partial causes of reionization, as they 
emit a significant numbers of ionizing photons\footnote{Here 
ionizing photons are those with an
energy of $E > 13.6 \unit{eV} = 2.18 \times 10^{-18} \unit{J}$, 
or a wavelength of 
$\lambda < 0.0912 \unit{\micron}$, 
sufficient to remove a ground state electron from a hydrogen atom.}.
While massive stars in ordinary galaxies were the dominant 
contributors to global 
reionization (\eg, \cite{Bouwens_etal:2012,Finkelstein_etal:2012}),
quasars can have a dramatic effect on their local environment,
being capable of 
creating Mpc-scale 
\hii\ regions\footnote{The \hii\ region formed by a quasar 
surrounded by a predominantly neutral IGM 
is similar to a classical Str\"{o}mgren sphere formed by 
an O or B star \cite{Stromgren:1939}.
The main difference is that a Str\"{o}mgren sphere is 
static, the continuous emission of ionizing radiation 
being balanced by recombinations in the inter-stellar medium,
whereas the density of the IGM during reionization is so
low that the \hii\ region around a high-redshift quasar 
can be expected to grow for the entirety of the quasar's lifetime.}
in an otherwise neutral IGM.
While the stars in galaxies 
can also produce \hii\ regions,
those produced by quasars
are considerably larger
than expected from an
individual galaxy \cite{Haiman:2002}
or even a proto-cluster \cite{Furlanetto_etal:2004}.
Towards the end of reionization the ionizing background from 
numerous more distant sources can be comparable to the radiation
even close to a quasar,
so the focus here is on \hii\ regions produced by quasars 
when reionization was still complete.

Assuming that the quasar is the 
dominant local source of (isotropically emitted) ionizing photons,
and that the surrounding IGM is uniform,
the (proper) radius of the spherical \hii\ region,
$\rhii$, evolves according to
\cite{Haiman:2002,Shapiro_Giroux:1987,Madau_etal:1999}
\begin{equation}
\label{equation:drdt}
\frac{\diff \rhii}{\diff t}
  = \frac{\rateion}{4 \pi \, \rhii^2 \, \nhi}
  + H \, \rhii
  - \nhi \, \alpha_{\rm B} \, \rhii,
\end{equation}
where 
$H$ is the (time-dependent) Hubble parameter,
$\rateion$ is the rate at which the quasar was emitting ionizing photons, 
and 
$\alpha_{\rm B} \simeq 2.6 \times 10^{-19} \unit{m}^3 \unit{s}^{-1}$ 
is the case B hydrogen recombination coefficient 
(with the value quoted appropriate for an IGM temperature of 
$\sim 10^4 \unit{K}$).
The first term on the right-hand side of \eq{drdt} 
corresponds to the production of ionizing photons by the quasar;
the second 
corresponds to the expansion of the \hii\ region with 
the Hubble flow, which can be ignored if the age of the quasar when
observed, $\age$, 
is much less
than the Hubble time (at the redshift in question); 
the third term corresponds to recombinations within the 
\hii\ region, and should be unimportant at redshifts of $\la 10$.
Making these approximations leads to the simple 
result that 
the $\rhii$ can be approximated by 
equating the number of ionizing photons that had been emitted 
by the quasar at the time it was observed\footnote{The 
fact that the ionization front
grows at an appreciable fraction of the speed of light
implies that care must be taken when calculating the
time that the quasar has been emitting ionizing radiation;
but the fact that the observed photons and the ionizing photons
take the same time to reach the edge of the NZ leads to
an exact cancellation in the case of the line-of-sight observations
discussed in \sect{nzobs} \cite{White_etal:2003,Yu_Lu:2005}.}
with the 
number of neutral hydrogen atoms in a volume 
$4 \pi \, \rhii^3 / 3$ of the IGM at that redshift,
which 
gives (\cf\ \cite{Haiman:2002,Madau_Rees:2000,Cen_Haiman:2000})
\begin{eqnarray}
\label{equation:rhii}
\rhii & = & 
  \left[ \frac{3 \, \age \, \rateion}
  {4 \pi \, \bar{n}_{\rm H,0} 
    \, (1 + \redshift)^3 \, \fhi(\redshift)} \right]^{1/3}
  \\
  & \simeq & 
  8.0 
  \left(\frac{\rateion}{6.5 \times 10^{57} \unit{s}^{-1}}\right)^{1/3}
   \!\!
  \left(\frac{\age}{2 \times 10^7 \unit{yr}} \right)^{1/3}
   \!\!
  \left(\frac{1 + \redshift}{7} \right)^{-1}
   \!\!
  \fhi^{-1/3} (\redshift)
  \unit{Mpc},
\nonumber \\
\end{eqnarray}
where \eq{nhicosmo} has been used to exhibit the dependence on
the neutral fraction
and $\redshift$ is the redshift of the quasar.
More realistic calculations of $\rhii$ 
have included effects 
such as departures from spherical symmetry,
over-densities inside the \hii\ region,
and 
fluctuations in quasar luminosity
(\eg, 
\cite{White_etal:2003,
Barkana_Loeb:2001,
Wyithe_Loeb:2004,
Wyithe_etal:2005,
Haiman_Cen:2005}),
but the scaling of $\rhii$ with the number of ionizing
photons and the local density of \hi\ is reasonably generic.

This leads to the appealing possibility of using the
measured \hii\ region sizes of \hzqs\ as a tracer of the
evolving neutral fraction,
although the various attempts to do so
(\eg,
\cite{
Yu_Lu:2005,
Wyithe_Loeb:2004,
Wyithe_etal:2005,
Mesinger_Haiman:2004,
Maselli_etal:2007,
Maselli_etal:2009,
Bolton_etal:2011,
Maselli_etal:2007})
have produced results that are at odds with other probes --
and, in some cases, contradict each other.
Most of these problems stem from the difficulty of estimating
$\rhii$ from line-of-sight absorption measurements (\sect{nzobs});
in the future 
more progress might be made 
with observations in the plane of the sky 
to obtain direct
images of \hzq\ \hii\ regions (\sect{nzimag}).


\subsection{Near zone measurements}
\label{section:nzobs}

\begin{figure}[t]
\begin{center}
\includegraphics[scale=.6]{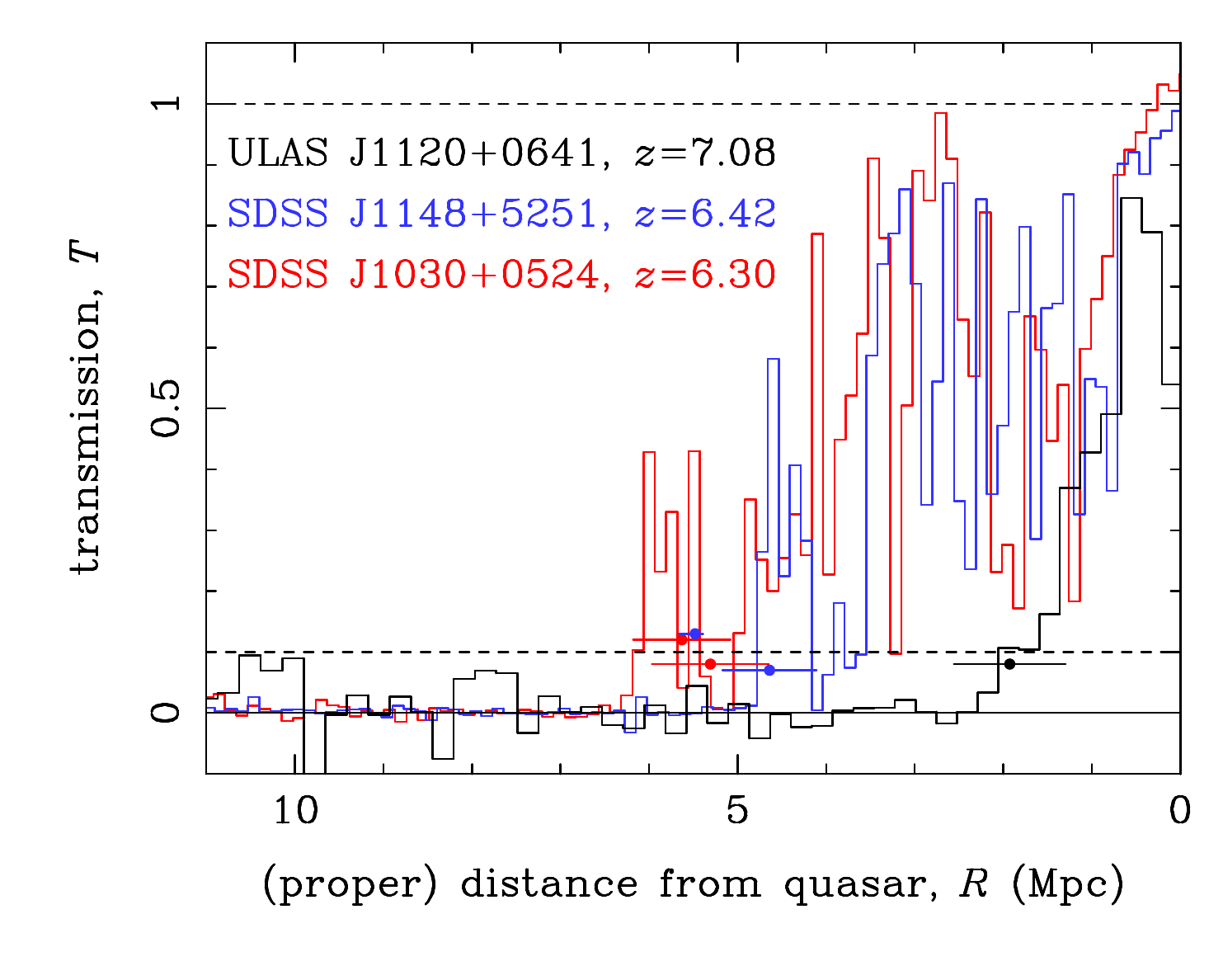}
\caption{The NZ transmission profiles of three \hzqs,
as labelled, showing both the systematic evolution with redshift
and the complicated \hi\ distribution along individual lines-of-sight.
The very different noise levels of the three spectra can be gauged 
from the fluctuations at distances of $R \ga 6 \unit{Mpc}$
as the IGM is essentially opaque, resulting in $T = 0$,
in front of these sources.
The dashed line at $T = 0.1$ corresponds to the value 
commonly used to estimate $\rnz$ 
(as described further in \sect{nzobs}).
The estimated $\rnz$ values,
calculated from the $\redshift_{\rm NZ}$ values
in \tabl{quasars}, are shown just below this line;
estimated $\rhii$ values \cite{Mesinger_Haiman:2007}
are shown just above this line.}
\end{center}
\label{figure:nzprofile}
\end{figure}

\begin{figure}[t]
\begin{center}
\includegraphics[scale=.6]{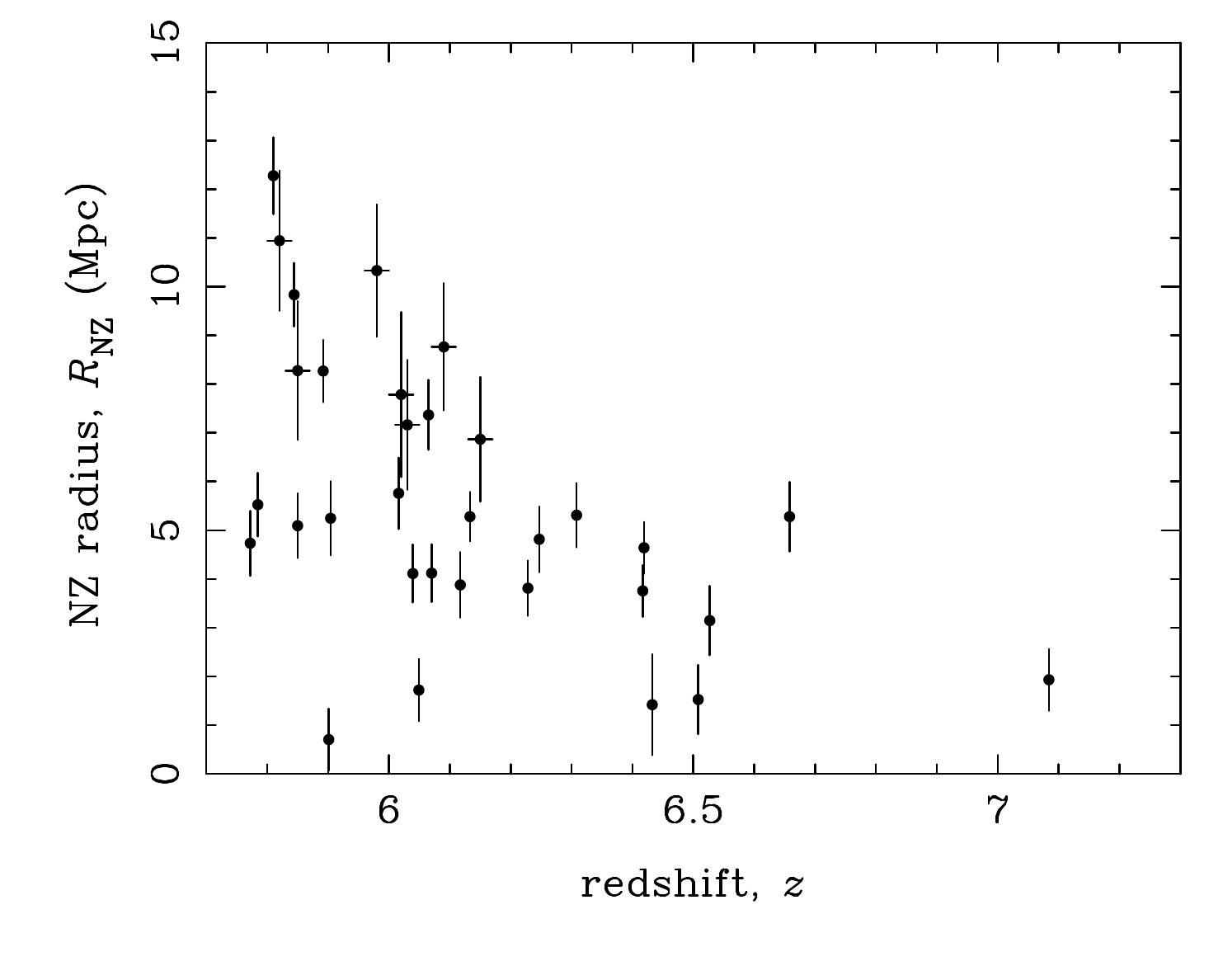}
\includegraphics[scale=.6]{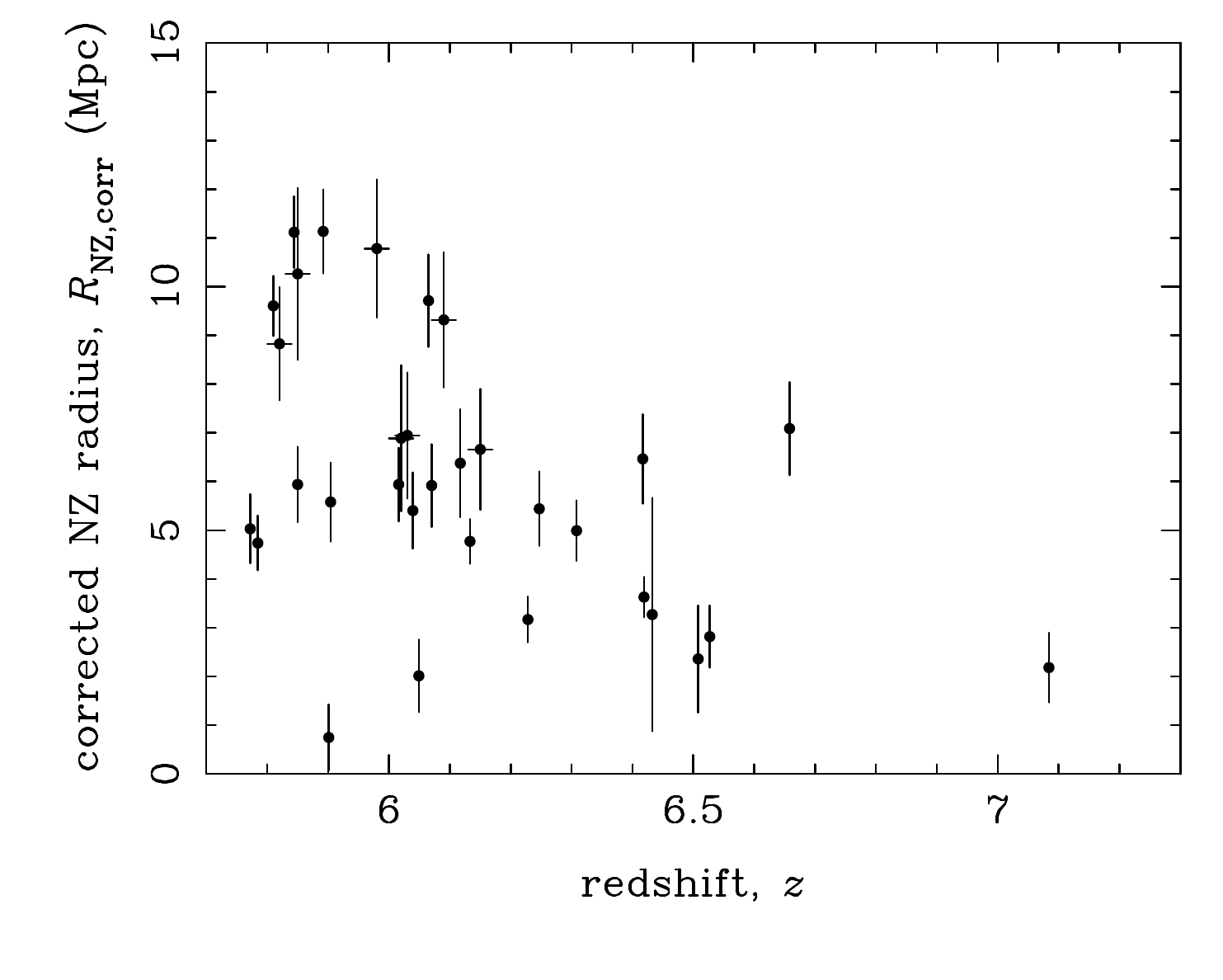}
\caption{Estimated proper radii (top panel) and corrected
  proper radii (bottom panel) of \hzq\ NZs,
  calculated using the fiducial cosmological model defined
  in \sect{cosmology} for the quasars
  with $\redshift_\nz$ measurements as listed in \tabl{quasars}.}
\label{figure:nearzones}
\end{center}
\end{figure}

The only currently practical option for measuring $\rhii$
around a \hzq\ is to exploit the fact that
the presence of \hii\ in front of a quasar results in
a near zone (NZ) of significant transmission
just blueward of each Lyman series emission 
line\footnote{The physical processes that produce a
quasar
NZ at $\redshift \ga 6$
are the same as those responsible for the proximity effect
(\eg, \cite{Bajtlik_etal:1988}) in lower redshift quasars.}.
Examples of observed NZ transmission profiles towards some \hzqs\
are shown in \fig{nzprofile}.
If the \hii\ region 
around a quasar at redshift $\redshift_\src$
was completely ionized then
there would be a region of complete transmission
just blueward of the $n$th Lyman series transition,
extending to a wavelength of (\cf\ \eq{lambda_r})
\begin{equation}
\label{equation:lambda_nz}
\lambda_{\obs,\subhii,n} \simeq
 \lambda_{1,n} \, (1 + \redshift_\src)
   \left[ 1 - \frac{\rhii}{c / H_0} 
    \Omega_{\rm m}^{1/2} (1 + \redshift_\src)^{3/2} \right].
\end{equation}
This wavelength is almost always observable for $n = 1$
(\ie, \lya)
and often for $n = 2$ (\ie, \lyb) as well,
but seldom for higher order transitions
as the relevant region of the spectrum is typically subject
to complete GP absorption (\sect{gp}) at these redshifts.

The most powerful approach to analysing the observed 
spectra in the wavelength range 
from $\lambda_{\obs,\subhii,n}$ to $\lambda_{1,n} \, (1 + \redshift_\src)$
is to use the transmission measurements to constrain a
physical
model of the 
residual \hi\ inside the \hii\ region
\cite{Mesinger_Haiman:2004,Mesinger_etal:2004,Bolton_Haehnelt:2007}.
In particular, such models have shown that the density of 
residual \hi\ 
towards the outside of the \hii\ region can be sufficent
to produce an optical depth of $\tau \gg 1$, 
meaning that the observable NZ of high transmission
would not not extend all the way to 
$\lambda_{\obs,\subhii,n}$ \cite{Mesinger_Haiman:2004,Bolton_Haehnelt:2007}.
But, as the 
effects of possible damped absorption from the IGM should also be included
in such an analysis,
a discussion of this is deferred to \sect{wing}.

The opposite approach,
which has the virtue of simplicity,
is to characterize the NZ in terms of 
the wavelength range of significantly positive transmission.
The specific definition most commonly used is 
to define $\lambda_{\obs,{\rm NZ},n}$ as
the observed wavelength at which 
the measured transmission,
binned on a reasonable but somewhat
arbitrary scale of $0.002 \unit{\micron}$,
first drops to 10\% of its peak value,
as measured bluewards from the $n$th Lyman series emission wavelength
\cite{Fan_etal:2006b}.
Converting to the
transition-independent NZ redshift,
$\redshift_\nz = \lambda_{\obs,\subhii,n} / \lambda_{1,n} - 1$,
then allows $\rnz$ to be estimated by 
inverting \eq{z_r} to give
\begin{equation}
\rnz \simeq
  \frac{c}{H_0}
   \frac{\redshift_\src - \redshift_{\rm NZ}}
   {\Omega_{\rm m}^{1/2} \, (1 + \redshift_\src)^{5/2}}.
\end{equation}
The observational uncertainty in 
$\redshift_\nz$ is
typically $\sigma_\nz \simeq 0.01$ \cite{Fan_etal:2006b,Carilli_etal:2010};
the uncertainty in the systemic source redshift, $\redshift_\src$,
depends primarily on how it has been estimated from the observed spectrum.
The least accurate, 
but most widely available,
option is to use 
the red wing of the \lya\ emission line, 
combined with the often blended \nv\ emission line
(rest-frame wavelength $\lambda = 0.1240 \unit{\micron}$),
for which $\sigma_\src \simeq 0.02$,
although the fact that this is not separate from the
spectral break is problematic
for making meaningful NZ measurements.
It is far more reliable and precise to
base systemic redshifts on the \mgii\ emission line
(rest-frame wavelength $\lambda = 0.2798 \unit{\micron}$),
which gives $\sigma_\src \simeq 0.007$
(\eg, \cite{Carilli_etal:2010,Hewett_Wild:2010}),
or the narrower mm-band molecular CO features,
which can yield $\sigma_\src \la 0.002$
(\eg, \cite{Carilli_etal:2010,Venemans_etal:2012a,Walter_etal:2003}).
By comparison, 
$\rhii \simeq 5 \unit{Mpc}$ at $\redshift_\src = 6$
gives $\redshift_\src - \redshift_\nz \simeq 0.1$,
implying that the resultant $\rnz$ measurements
are at least statistically meaningful --
but this does not imply that $\rnz$ is a good proxy for $\rhii$.

All the published measurements
of $\redshift_\src$ and $\redshift_\nz$
for $\redshift_\src \geq 5.8$ quasars
are tabulated in \tabl{quasars}.
The resultant estimates of 
both the actual NZ radius, $\rnz$,
and the luminosity-corrected NZ radius\footnote{Equation~\ref{equation:rhii}
makes it clear that 
estimates for both
$\rateion$ and $\age$
are needed if any attempt is to be made to infer
$\fhi$ from the scale of either the NZ or the \hii\ region.
It is at least plausible to assume that the ionization rate is
proportional to
the quasar's luminosity,
$L$, at the moment of observation, leading to the use of
the ``corrected'' NZ radius \cite{Fan_etal:2006b}
$\rnzcorr = \rnz \, 10^{2/5[\mabs - (-27)] / 3} \propto \rnz / L^{1/3}$,
defined so that quasars of different luminosities can be compared
on an approximately equal footing.},
$\rnzcorr$, are 
plotted against 
$\redshift_\src$ in \fig{nearzones}.
A clear anti-correlation is apparent,
implying that 
the IGM 
was in a significantly different state 
at $\redshift \simeq 6$ than it had been
$\sim 10^8 \unit{yr}$ earlier at $\redshift \simeq 7$.
But it is difficult to
make more quantitive statements, due 
in part to the scatter that 
results from 
the range of 
intrinsic quasar properties 
(mainly age at time of observation)
and environment 
(the density and distribution of hydrogen in the surrounding IGM),
but largely because $\rnz$ is significantly lower than
$\rhii$ in general.
It is hence unlikely that \hzq\ NZ measurements will prove
to be the accurate probe of the cosmological reionization history
that was initially hoped for,
although they will continue to be measured as 
a natural fringe benefit of 
obtaining accurate systemic \hzq\ redshifts.


\subsection{Imaging}
\label{section:nzimag}

A strong validation 
of the overall reionization paradigm 
would be if an \hzq\ \hii\ region could be imaged directly,
rather than just inferred from a spectroscopic absorption profile.
An \hii\ region of (proper) radius $\sim 5 \unit{Mpc}$ at $\redshift = 6$
would have an angular radius of $\sim 0.2 \unit{deg}$, 
resolvable using current instruments at most wavelengths.
The choice of wavelength is key, however, as it there must be 
some emission (or absorption) signature by which the NZ is to 
be distinguished from its surroundings.

One promising possibility is collisional excitation of 
the residual \hi\ in the \hii\ region
which could be seen in an appropriately tuned filter
\cite{Cantalupo_etal:2008}.
This effect is stronger with a higher neutral fraction,
and so the most promising targets for observation 
are the most distant known quasars -- 
a High Acuity Wide field K-band Imager (HAWK-I \cite{Kissler-Patig_etal:2008})
program to observe ULAS~J1120+0641 at $\redshift = 7.08$ 
is currently underway.

In the longer term, radio observations of redshifted 21~cm 
emission from -- and absorption by -- \hi\ should yield direct
images of large numbers of \hzq\ \hii\ regions
\cite{Wyithe_etal:2005,Madau_etal:1997}.


\section{The \lya\ damping wing}
\label{section:wing}

\begin{figure}[t]
\begin{center}
\includegraphics[scale=.6]{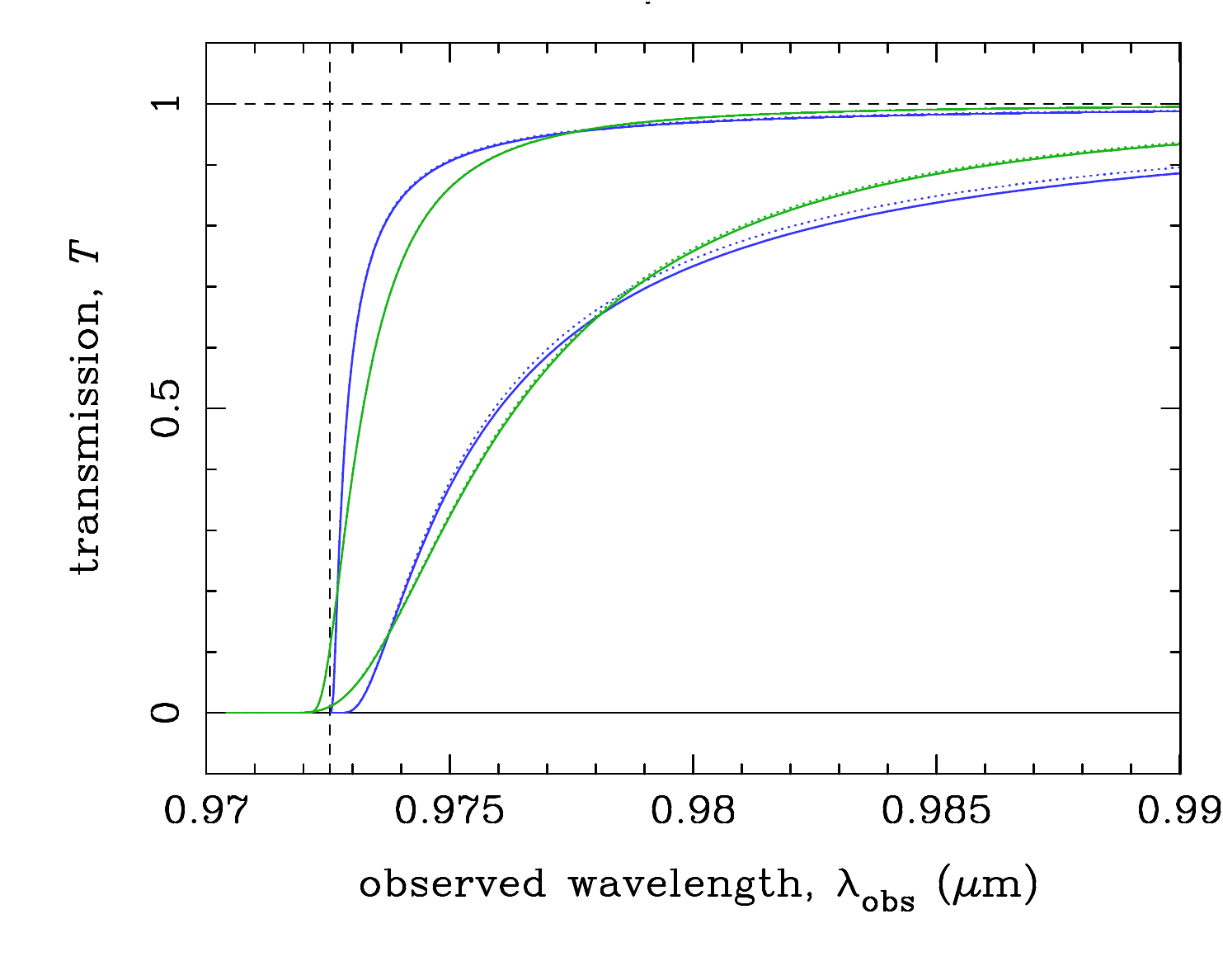}
\caption{Damped \lya\ transmission 
  profiles as produced by extended IGM absorption (blue)
  and a DLA (green).  
  The IGM profiles
  are for the default cosmological parameters in \sect{cosmology}
  with constant neutral fractions of $\fhi = 0.1$ (top)
  and $\fhi = 1.0$ (bottom)
  between $\redshift_\start = 7.0$ and $\redshift_\finish = 6.0$.
  The vertical dashed line indicates the wavelength of 
  \lya\ at $\redshift = 7.0$.
  The DLA profiles are matched to these, 
  with 
  $\Sigma_\subhi = 6.0\times10^{23} \unit{m}^{-2}$
  at $\redshift_\abs = 6.993$ for the $\fhi = 0.1$ model
  and 
  $\Sigma_\subhi = 1.0\times10^{25} \unit{m}^{-2}$
  at $\redshift_\abs = 6.980$ for the $\fhi = 1.0$ model.
  All curves are paired, showing the 
  results for the Lorentzian line profile (solid)
  and the two-level line profile (dotted); 
  these are not always distinguishable.}
\end{center}
\label{figure:dampingwing}
\end{figure}

If the IGM in front of a 
source was mostly neutral then there
would be appreciable absorption 
redward of the wavelength of \lya\ photons leaving the 
quasar's \hii\ region (\sect{nearzone}),
and possibly reward
of the systemic \lya\ emission wavelength as well.
This comes about due to the sum of the 
Rayleigh scattering wings,
which combine to give the long wavelength damped absorption 
seen clearly in Figs~\ref{figure:crosssection} and \ref{figure:dampingwing}.

The mere detection of such a damping wing would,
provided it could be shown that it did not arise 
from a discrete concentration of \hi,
be direct evidence that $\fhi \ga 0.1$.
Further, as the depth and extent of the damping wing 
are determined by the density of \hi\ in front of the source,
the observed transmission profile
could be used to measure $\fhi$ in front of the source.
The fact that the damping wing 
is many orders of magnitude weaker 
than the resonant absorption that produces the GP
effect (\sect{gp}) makes it
a potentially useful probe of redshifts
at which the hydrogen in the IGM is completely neutral.
Conversely, 
there is also information to be gleaned from the absence of 
a detectable damping wing, as this places an upper limit on 
the \hi\ density in front of a source.
Any such constraint is robust to the presence of 
a discrete 
\hi\ concentration:
the upper limit on $\fhi$ could only be made stronger as a result.


\subsection{IGM absorption profile}
\label{section:igm_profile}

\begin{figure}[t]
\begin{center}
\includegraphics[scale=.6]{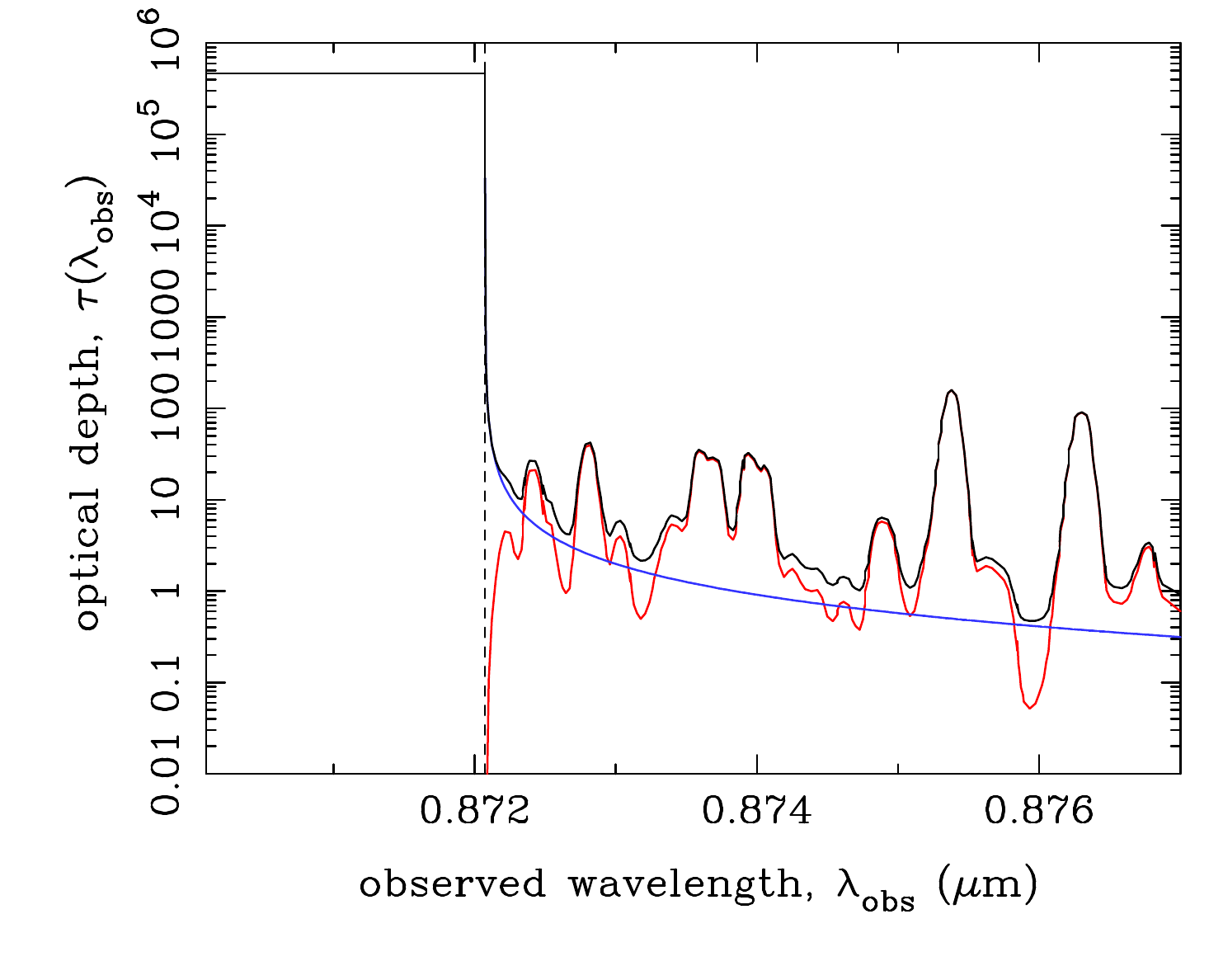}
\caption{The optical depth along a typical line-of-sight
towards a
quasar at $\redshift = 6.28$ (such as SDSS~J1030+0524).  
The quasar is assumed to be embedded in a fully
neutral, smooth IGM, but surrounded by a \hii\ region with a 
(proper) radius
of $R = 6 \unit{Mpc}$.  
The red curve shows the contributionfrom residual \hi\ inside 
the \hii\ region;
the blue curve shows the contribution 
from \hi\ outside the \hii\ region in the surrounding IGM;
the black curve shows 
the total optical depth,
given by the sum of these two contributions.
The vertical dashed line indicates the 
edge of the \hii\ region.
For reference, the
redshifted \lya\ emission wavelength is at $0.8852 \unit{\micron}$, 
far to the right off the plot. 
Adapted from \cite{Mesinger_Haiman:2004}.}
\end{center}
\label{figure:tau_regions}
\end{figure}

The canonical IGM damping profile is calculated by adopting the 
simple model that $\fhi$ was constant between the front of 
any \hii\ region around the source (\sect{nearzone}), 
at redshift $\redshift_\start$,
and the hard end-point of reionization, at 
redshift $\redshift_\finish$. 
Inserting this simple reionization history into
\eq{tau_general} gives the optical depth at an observed wavelength 
of $\lambda_\obs > (1 + \redshift_\start) \, \lambda_\alpha$ 
as (\cf\ \cite{Miralda-Escude:1998,Miralda-Escude_Rees:1998})
\begin{equation}
\label{equation:wing}
\tau(\lambda_\obs)
  = 
    \frac{c \, \bar{n}_{\rm H,0} \, \sigma_{\rm T} \, f_{1,2}^2 \, \fhi}
      {\Omega_{\rm m}^{1/2} \, H_0}
    \left( \frac{\lambda_\obs}{\lambda_\alpha} \right)^{3/2}
   \left\{
     I\left[
\frac{(1 + \redshift_\start) \lambda_\alpha}{\lambda_\obs}
     \right]
     -
     I\left[
\frac{(1 + \redshift_\finish) \lambda_\alpha}{\lambda_\obs}
     \right]
   \right\},
\end{equation}
where 
\begin{equation}
I(x) 
  = \int \diff x \, x^{1/2} \, \frac{\sigma(x \, \nu_\alpha)}
    {\sigma_{\rm T} \, f_{1,2}^2}
\end{equation}
is the dimensionless integral that 
results from changing 
the integration variable from 
$\redshift$
to $x = \nu / \nu_\alpha
  = (1 + \redshift) \, \lambda_\alpha / \lambda_\obs$.
In contrast to GP absorption
(\sect{gp}),
there is no meaningful conversion from $\lambda_\obs$ to $\redshift$,
as the flux received at any given wavelength
is gradually attenuated as it passes through the IGM.

The strength of the absorption at the redshifted \lya\ 
wavelength can be gauged by comparison to the GP
optical depth given in \eq{tau_gp}: 
identifying $\lambda_\obs / \lambda_\alpha$ as $\redshift_\src$,
the pre-factor in \eq{wing} can be re-written as 
$2 \, \einstein_{1,2} \, \lambda_\alpha /(3\, c)
  \, \tau_{\gp,\alpha}(\redshift_\src)
\simeq 10^{-7} \, \tau_{\gp,\alpha}(\redshift_\src)$.
It is this comparative weakness of the damping wing absorption 
that makes it a potentially useful probe of a largely neutral IGM.

If the full cross section from \eq{crosssection_full}
is used then the integration must be evaluated numerically,
but a good combination of simplicity and accuracy 
($\sim 1\%$ relative error) can be
achieved by using the simple Lorentzian
form of $\sigma(\nu)$ given in \eq{crosssection_lorentzian_nonres}.
This yields 
\begin{equation}
I_{\rm Lor}(x) \simeq 
  \int \diff x \, \frac{x^{1/2}}{4 \, (x - 1)^2} 
  = \frac{x^{1/2}}{4 \,(1 - x)} + 
  \frac{1}{2} \, \ln \left( \frac{1 - x^{1/2}}{1 + x^{1/2}} \right)
\end{equation}
and gives the IGM damping wing profiles shown as the 
solid curves in \fig{dampingwing}. 
Using the two-level model from \eq{crosssection_miralda} 
gives the widely used result that (\cf\ \cite{Miralda-Escude:1998})
\begin{eqnarray}
I_{\rm 2L}(x) 
  & 
  \simeq 
  &
  \int \diff x \frac{x^{9/2}}{4 \, (x - 1)^2}
\nonumber
\\
 &
 =
 & 
  \frac{
  x^{1/2} \, (315 - 210 \, x - 42 \, x^2 - 18 \, x^3 - 10 \, x^4)
  }{140 \, (1 - x)}
  +\frac{9}{8} \, \ln \left( \frac{1 - x^{1/2}}{1 + x^{1/2}} \right),
\end{eqnarray}
shown as 
the dotted curves in \fig{dampingwing}.
Even though the cross section of the two-level model is
too low by a factor of $\sim 10$ in the long-wavelength limit
(\sect{lyman_series}), most of the absorption occurs when the photons
first enter the IGM,
the regime in which the discrepancy is smallest.
Hence the two-level IGM transmission profile under-estimates 
the absorption by $\la 5\%$,
although the Lorentzian --
or a more accurate approximation
\cite{Bach_Lee:2015,Mortlock_Hirata:2015}
--
should be used in its place.

While it is an important point of principle to use 
an appropriate form of the wavelength-dependent cross section
in any calculation 
of the IGM damping wing, 
there are several limitations of
the simple model above that can have at least as much impact on
the transmission profile:

\begin{itemize}

\item 
The IGM is not uniform in density,
and so the clumped residual \hi\ inside the \hii\ region 
around a quasar can produce 
a transmission signature like that of the \lya\ forest
\cite{
Mesinger_Haiman:2007,
Maselli_etal:2007,
Mesinger_etal:2004,
Bolton_Haehnelt:2007,
Maselli_etal:2004}.

\item 
The \hii\ region of a quasar will contain other ionizing sources
(\ie, galaxies), 
which contribute to the local ionization balance. 
The relative importance of these galaxies
increases with distance from the quasar, 
and is hence greatest close to the edge of the \hii\ region
(\eg, \cite{Bolton_Haehnelt:2007b}).

\item Quasars are expected to be biased,
and so will reside in overdensities of galaxies and gas
on a scale of a few Mpc
\cite{Yu_Lu:2005,Wyithe_Loeb:2007,Wyithe_etal:2008}.  
The effects of increased ioinizing radiation 
would tend to be cancelled out by 
the extra gas, and so detailed simulations are required 
to assess whether the net effect is significant.

\item 
The interior of the \hii\ region can become optically thick to
ionizing radiation, necessitating a full
radiative transfer calculation
(\eg, the quasar's flux will drop more rapidly than with the 
square of distance from the quasar
\cite{Kramer_Haiman:2009,Thomas_Zaroubi:2008}).  Describing the
distribution of optically thick clumps (\ie, ``Lyman limit systems'')
near the quasar is one of the largest, and most challenging, modeling
uncertainties \cite{Crociani_etal:2011}.

\item 
In addition to the nonuniform density, the IGM outside the \hii\ 
region is ionized in patches, rather than uniformly \cite{Haiman:2011},
which will 
make the wavelength-dependence of the damping wing 
differ from the above uniform IGM model 
\cite{Bolton_Haehnelt:2007,Lidz:2007,McQuinn_etal:2008,Schroeder_etal:2013}.
Simulations indicate that there is significant bias in the 
inferred reionization parameters if this effect is not taken into
account \cite{Mesinger_Furlanetto:2008,Mesinger_etal:2015}.

\end{itemize}

One approach to dealing with these issues is to 
use numerical simulations of the \hi\ distribution around a \hzq\
(\eg, \cite{Mesinger_Haiman:2004,Mesinger_etal:2004,Bolton_Haehnelt:2007}).
The results of this approach are illustrated in \fig{tau_regions}, 
showing the separate contributions of the \hi\ inside and outside 
the \hii\ region towards a fiducial \hzq.
As can be seen, the 
contributions from the two distinct components are significant;
both can be accounted for in a pixel optical depth analysis as
discussed further in \sect{obs_profile} below.


\subsection{DLA profile}
\label{section:dla_profile}

A high column density \hi\ cloud in front of a source would 
also produce damped absorption that, given realistic observational
uncertainties, could be effectively indistinguishable from 
an IGM damping wing. 
Assuming the local IGM has $\fhi \ga 10^{-4}$, the 
GP absorption blueward of the NZ edge would leave just the 
red damping wing 
(as distinct from a DLA in an ionized medium, for which
both wings of the \lya\ line are often seen).
The proper density of an
a thin \hi\ cloud of column density $\Sigma_\subhi$ 
at a redshift of $\redshift_\abs$
can be approximated as
\begin{equation}
\nhi(\redshift) = 
  \frac{(1 + \redshift) \, H(\redshift) \, \Sigma_\subhi}{c}
  \diracdelta(\redshift - \redshift_\abs),
\end{equation}
where the line element from \eq{dl} has been used 
to change the argument of the delta function from the local
line-of-sight spatial coordinate to redshift.
Inserting this into \eq{tau_highz}
then gives the standard
DLA optical depth
\begin{equation}
\label{equation:dla}
\tau(\lambda_\obs) = \Sigma_\subhi \,
  \sigma_{\rm eff} \! 
  \left( \frac{1 + \redshift_\abs}{c / \lambda_\obs} \right),
\end{equation}
where $\sigma_{\rm eff}(\nu)$ is the velocity-convolved 
effective cross section defined in \eq{sigma_eff}.
If line-of-sight velocities are unimportant 
(\eg, significantly redward of resonance) then 
the Lorentzian cross section from
\eq{crosssection_lorentzian_nonres} can be used to give
\begin{equation}
\tau(\lambda_\obs) \simeq 
  \frac{ \Sigma_\subhi \, \sigma_{\rm T} \, f_{1,2}^2} 
  {4 [(1 + \redshift_\abs) \, \lambda_\alpha / \lambda_\obs - 1]^2},
\end{equation}
which is compared to the IGM transmission profiles in \fig{dampingwing}.
(If velocities are important then the more complicated Voigt
profile given in \sect{expand} should be used.)

The similarity between the tuned 
DLA profile and the IGM profiles shown in \fig{dampingwing}
is such that it will require exquisite quality spectra
to distinguish between them on the basis of data alone
\cite{Mortlock_Hirata:2015,Miralda-Escude:1998,McQuinn_etal:2008,
 Patel_etal:2010},
although there are other approaches to tackling this problem,
as discussed below.


\subsection{Observational issues}
\label{section:obs_profile}

\begin{figure}[t]
\begin{center}
\includegraphics[scale=.6]{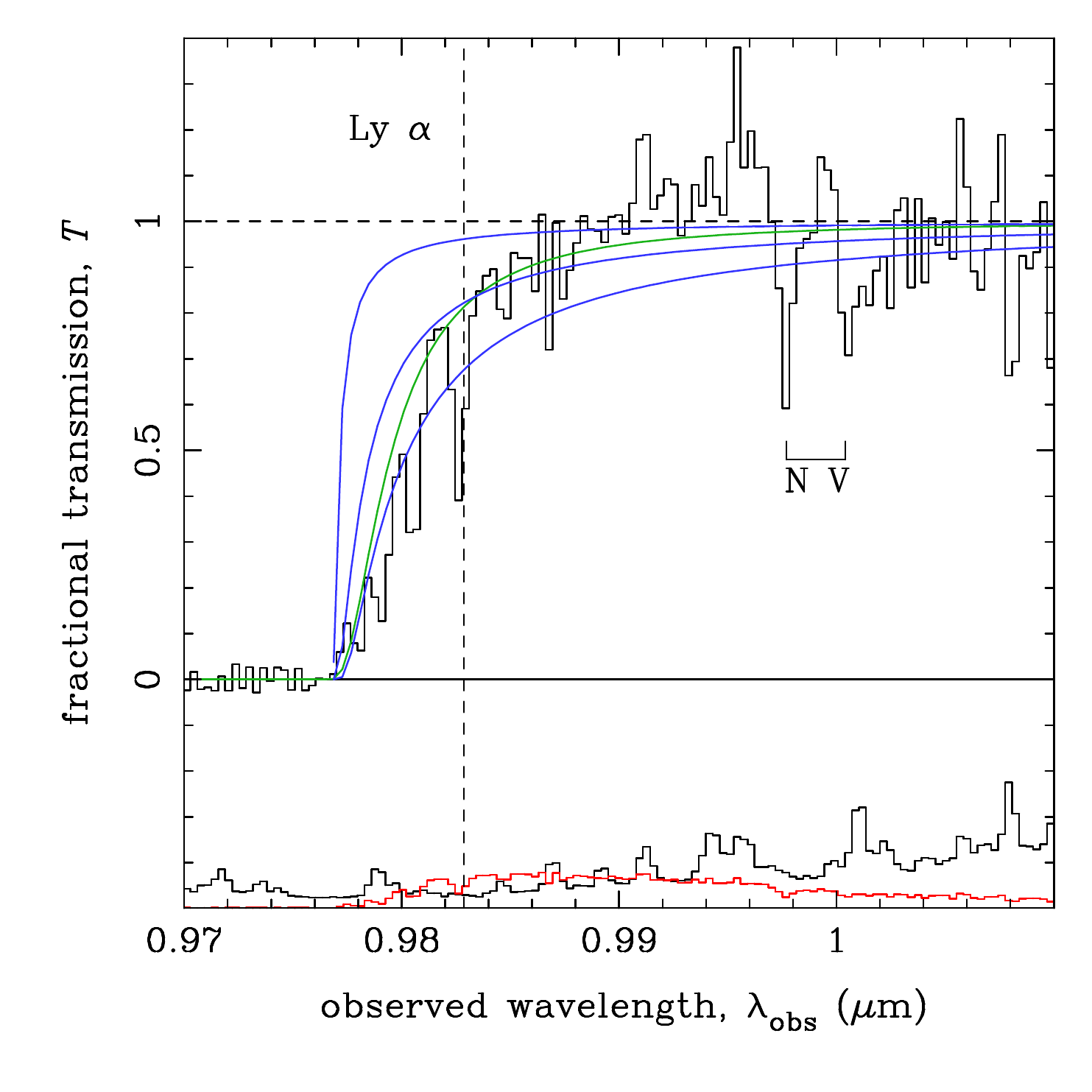}
\caption{Estimated transmission towards
  the $\redshift = 7.08$ quasar
  ULAS~J1120+0641 obtained by dividing the
  observed spectrum by a model for its intrinsic
  emission based on lower redshift quasar spectra
  \cite{Mortlock_etal:2011d}.  
  The resultant systematic uncertainty is shown at the bottom in 
  red, along with the observational noise in black.
  Shown in blue are theoretical IGM damping wing profiles
  for the default cosmological parameters in \sect{cosmology}
  with constant neutral fractions of $\fhi = 0.1$,
  $\fhi = 0.5$ and $\fhi = 1.0$ (from top to bottom)
  between $\redshift_\start = 7.035$ and $\redshift_\finish = 6.0$.
  A DLA profile is shown in green for an \hi\ cloud of column density
  $\Sigma_\subhi = 4\times10^{24} \unit{m}^{-2}$
  at $\redshift_\abs = 7.025$. 
  (These curves differ slightly from those shown in 
  \cite{Mortlock_etal:2011d} due to the use of the Lorentzian
  line profile and the inclusion of a helium fraction of 
  $Y = 0.24$.)}
\end{center}
\label{figure:dampingwing_1120}
\end{figure}

The challenges of identifying and interpreting damped IGM 
absorption towards a distant quasar are not limited to the theoretical
issues described above: 
there are significant observational challenges as well,
both blueward and redward of the quasar's systemic \lya\ wavelength.

Any damped absorption would be strongest 
at the front edge of the \hii\ region 
around the quasar; and, 
assuming a canonical (proper) radius of 
$\sim 5 \unit{Mpc}$ (\sect{nearzone}),
it is only blueward of the systemic \lya\ 
wavelength that an IGM damping wing would be visible
if $\fhi \la 0.1$,
and it is hence in this wavelength range
that searches for damped absorption 
towards $\redshift \la 6.5$ quasars have concentrated to date.
The 
main complication is the need to model 
the effects of the residual \hi\ inside the quasar's \hii\ region,
as discussed in \sect{igm_profile},
with uncertainty in 
the quasar's intrinsic emission 
a secondary issue \cite{Kramer_Haiman:2009}.
The results of these pixel optical depth
studies are inevitably strongly model-dependent,
although the
fact that the
residual \hi\ should also show the same transmission pattern
blueward of the \lyb\ line can provide an important check
\cite{Mesinger_Haiman:2004}.
Applying this approach to three bright $6.2 < \redshift < 6.5$
SDSS quasars implied that
$\fhi \simeq 1$ at $\redshift \simeq 6$ along two of these lines-of-sight
\cite{Mesinger_Haiman:2007};
a subsequent improved analysis \cite{Schroeder_etal:2013}
updated these constraints to be $\fhi \ga 0.1$.

The most unambiguous detection of
damped absorption would be 
if it could be confirmed redward of of the quasar's
systemic \lya\ wavelength,
as it is only the damping wing absorption from cosmological \hi\ 
patches 
(and/or a possible a DLA) that need be taken into account --
the (uncertain) distribution of residual \hi\ 
would not have a significant effect and could be ignored.
The main difficulty is that an estimate of the fractional transmission --
and particularly its wavelength dependence -- 
requires an accurate model 
of the unabsorbed spectral energy distribution of the quasar,
specifically the strong \lya+\nv\ emission lines\footnote{This is 
not an issue with GRBs, 
which have very smooth spectra at these wavelengths
(\eg, \cite{Patel_etal:2010,Totani_etal:2006,Chornock_etal:2015}).}.
Plausible results can be obtained by fitting a paramterised
model spectrum (\eg, \cite{Mesinger_Haiman:2004}),
but the fact that the properties of quasars' \lya+\nv\ lines are correlated
with the unabsorbed emission at longer wavelengths
(\eg, \cite{Hewett_Wild:2010}) 
should be exploited if possible.

This approach has been applied to the 
$\redshift = 7.08$ quasar ULAS~J1120+0641,
which does appear to exhibit absorption redward of \lya\
\cite{Mortlock_etal:2011d},
as shown in \fig{dampingwing_1120},
although this view is not unanimously held
\cite{Bosman_Becker:2015}.
Despite significant systematic uncertainties
from the inevitable ignorance about the quasar's true emission,
the identification of absorption redward of \lya\ 
is supported by completely independent data
and analysis methodology \cite{Simcoe_etal:2012}.
There is, however,
the remaining question of whether the 
absorption is due to the IGM
(which would imply $\fhi \ga 0.1$ at $\redshift \simeq 7$)
or a DLA
(with a very high column density of 
   $\Sigma_\subhi \simeq 10^{24} \unit{m}^{-2}$).
Simulations of the \hi\ distribution in such
systems imply that such isolated concentrations
are rare, being expected along only one line-of-sight in $\sim 20$
\cite{Bolton_etal:2011};
and,
given also that numerical simulations imply such pristine gas
would not last to that epoch \cite{Finlator_etal:2013},
the lack of any associated metal
absorption lines \cite{Simcoe_etal:2012}
also disfavours the DLA hypothesis.
It is also tempting to try and interpret the 
shape of the damped profile,
although to do this correctly will require
that the impact of patchy ionization in the IGM 
and residual \hi\ in the \hii\ region be accounted for.

An obvious approach to verifying and extending these
results would be to apply the same analysis to the 
more recently discovered
$\redshift \ga 6.5$ quasars
\cite{Venemans_etal:2015,Venemans_etal:2013}.
These sources 
have not yet been analysed in great detail,
but 
the necessary high signal-to-noise ratio spectroscopic observations will
presumably be made in due course.
Another way of making progress here,
particularly on the IGM vs.\ DLA question,  would
be by finding more bright $\redshift \simeq 7$ quasars:
if damping wings were rare then the implication
would be that the IGM was fairly ionized at this epoch;
if damping wings were ubiquitous then it
would be unequivocal that the IGM was significantly neutral.


\section{Future observational prospects}
\label{section:future}

A recurring theme above is that more bright quasars with 
redshifts of $\redshift \ga 6$ need to be identified
to establish the global evolution of the hydogren in the IGM.
In terms of numbers, the Dark Energy Survey (DES \cite{DES:2005})
and Pan-STARRS should soon eclipse SDSS and the CFHQS,
after which 
the
Large Synoptic Survey Telescope (LSST \cite{Ivezic_etal:2008})
will, from its first scans, be able to complete the census of
quasars with $\redshift \la 6.5$.
All these surveys only extend to the $Y$ band,
which will limit their ability to explore higher redshifts
without complementary NIR data.

The fact that there are currently only seven confirmed
optical drop-out quasars
with $\redshift \ga 6.5$ means that there is still considerable
amount to be done with NIR-based quasar searches,
to at least provide a sample that can be compared with the
optically-selected $\redshift \la 6.5$ objects.
Unfortunately, this is likely to be considerably more
challenging than merely
``repeating'' SDSS-like searches at longer wavelengths:
the quasar luminosity function decreases with redshift
as $\sim 10^{-0.5 \redshift}$
\cite{Fan_etal:2001}
and so the level of contamination by Galactic
stars and brown dwarfs will increase accordingly.
Still, progress will be made here, even if slowly
\cite{Willott_etal:2010}:
two $\redshift \ga 6.5$ quasars have been detected in UKIDSS;
and
VISTA will yield more sources at comparable distances.
Both surveys have sufficient wavelength coverage that they could
also probe $\redshift \ga 8$,
although their area coverage is such that it is possible
that neither will yield such $Y$-band drop-out quasars.
Further ahead, Euclid \cite{Laureijs_etal:2011}
should revolutionise this field
due to its large ($\sim 20,000 \unit{deg}^2$) area coverage
and comparatively deep NIR imaging.
There is also the possibility that Euclid will find 
\hzqs\ using low-resolution spectroscopy \cite{Roche_etal:2012}.

Finding new and more distant quasars with these surveys will,
however, only be useful for probing reionization if it is possible
to obtain deep spectroscopic observations of these objects,
so there is a trade-off between area and depth.
Given that the ultimate observational aim --
to have accurate absorption measurements along many lines-of-sight -- 
is essentially the same, which of these two options can be most
usefully pursued depends in part upon the resources available.
For instance, both the Thirty Metre Telescope (TMT)
and Giant Magellan Telescope (GMT)
ought to be able to obtain spectra
of all the known \hzqs\ that are comparable in quality to those
which are currently available only for, \eg, 
SDSS~J1148+0521 \cite{White_etal:2003}.
Conversely, the decreased pressure on smaller telescopes 
like those used for the SDSS and UKIDSS projects 
might make it possible to greatly increase the
area covered by these types of surveys, potentially
yielding \hzqs\ a magnitude or more brighter than those known at present.

The next decade should also see a new paradigm for exploring the 
reionization epoch, in the form of 21~cm observations of 
neutral hydrogen at high redshifts (Chapter~X).
The Square Kilometre Array (SKA)
and the various ``pathfinder'' projects 
will be able to both detect \hzqs\
\cite{Wyithe_etal:2005b} and 
characterise, in particular, their \hii\ regions
\cite{Wyithe_etal:2005,Madau_etal:1997,Alvarez_Abel:2007,Wyithe_etal:2015}.

Such data would go a long way to providing a full empirical 
characterisation of hydrogen in the reionization epoch, 
although real understanding will only come in the context of 
theoretical models.  
The process of reionization is sufficiently complicated that 
progress here will inevitably come from numerical simulations.  
At present the available simulations are either fast and global
but approximate (\eg, \cite{Greig_etal:2011,Mesinger_etal:2011})
or more complete but limited to small volumes 
(\eg, \cite{Finlator_etal:2013,Cantalupo_Porciani:2011,Iliev_etal:2014}). 
Inevitable increases in computing power will ensure steady 
progress, as will the increasingly tight observational constraints
that will obviate the need for such a wide range of initial 
conditions to be considered.  

The final step will be bringing observation and theory together,
something which has largely been tackled using fairly heuristic methods
so far.  While somewhat understandable given both the ambiguities
in the observational signatures and the large number of unknown
parameters in the models, it should become possible to deploy
more rigorous methods in the future
(\eg, \cite{Greig_Mesinger:2015}).  
One apparently fundamental difficulty of using numerical models
is that it is often impossible to write down a useful likelihood function --
simply comparing, \eg, the transmission along the line-of-sight to a quasar
with one realisation from a numerical simulation will almost always yield
a bad fit.
One promising option in such situations is approximate Bayesian computation
(ABC; \eg, 
  \cite{Tavare_etal:1997,Pritchard_etal:1999,Cameron_Pettitt:2012}), 
in which it is sufficient to be able to simulate mock data.
In practice it requires the use of cleverly chosen summary statistics
(\eg, the mean redshifts and lengths of the dark gaps discussed in 
\sect{gaps}), but as long as some information is encoded then 
rigorous inferences can be made.  
There is reason to be hopeful that quasar studies of reionization will,
over the next decade, evolve from their current exploratory nature
to provide rigorous quantitative constraints on the progress of 
reionization in the early Universe.


\begin{acknowledgement}
Thanks to 
Xiahoui Fan, 
Zoltan Haiman,
Chris Hirata,
Linhau Jiang,
Leon Lucy,
Andrei Mesinger, 
Subu Mohanty,
Ashara Peiris, 
Andrew Pontzen,
Steve Warren
and 
Chris Willott 
for useful discussions about quasars, reionization and 
the rich physics of the hydrogen atom.
\end{acknowledgement}


\bibliographystyle{unsrt.bst}
\bibliography{references}


\end{document}